\documentclass[10pt,twocolumn,amsmath,amssymb,aps,pra,secnumarabic,%
    nofootinbib,groupedaddress]{revtex4-1}
\usepackage{times,mathdots,mathptmx,amsthm,tikz}
\newtheorem{theorem}{Theorem}[section]
\newtheorem{proposition}[theorem]{Proposition}
\newtheorem{lemma}[theorem]{Lemma}
\newtheorem{corollary}[theorem]{Corollary}
\theoremstyle{remark}
\newtheorem*{remark}{Remark}

\newcommand{\R}{\mathbb{R}}
\newcommand{\Q}{\mathbb{Q}}
\newcommand{\Z}{\mathbb{Z}}
\newcommand{\C}{\mathbb{C}}
\newcommand{\eps}{\epsilon}

\newcommand{\AnPP}{\mathsf{A_0PP}}
\newcommand{\BPP}{\mathsf{BPP}}
\newcommand{\BQP}{\mathsf{BQP}}
\newcommand{\NP}{\mathsf{NP}}
\newcommand{\coNP}{\mathsf{coNP}}
\newcommand{\FP}{\mathsf{FP}}
\newcommand{\RP}{\mathsf{RP}}
\newcommand{\PH}{\mathsf{PH}}
\newcommand{\PP}{\mathsf{PP}}
\newcommand{\APX}{\mathsf{APX}}
\newcommand{\APV}{\mathsf{APV}}
\newcommand{\ARV}{\mathsf{ARV}}
\newcommand{\PTAS}{\mathsf{PTAS}}
\newcommand{\FPTAS}{\mathsf{FPTAS}}
\newcommand{\FPTEAS}{\mathsf{FPTEAS}}
\newcommand{\FPRAS}{\mathsf{FPRAS}}
\newcommand{\PostBPP}{\mathsf{PostBPP}}
\newcommand{\PostBQP}{\mathsf{PostBQP}}
\newcommand{\poly}{\mathrm{poly}}
\renewcommand{\P}{\mathsf{P}}
\newcommand{\SBP}{\mathsf{SBP}}
\newcommand{\SBQP}{\mathsf{SBQP}}
\newcommand{\PromiseSBP}{\mathsf{PromiseSBP}}
\newcommand{\PromiseSBQP}{\mathsf{PromiseSBQP}}
\newcommand{\PromisePostBPP}{\mathsf{PromisePostBPP}}
\newcommand{\PromisePostBQP}{\mathsf{PromisePostBQP}}
\newcommand{\shP}{\mathsf{\#P}}

\newcommand{\longto}{\longrightarrow}
\newcommand{\PSU}{\mathrm{PSU}}
\newcommand{\PSL}{\mathrm{PSL}}

\newcommand{\yes}{\mathrm{yes}}
\newcommand{\no}{\mathrm{no}}
\newcommand{\retry}{\mathrm{retry}}
\newcommand{\vy}{{\vec{y}}}
\newcommand{\cC}{\mathcal{C}}
\renewcommand{\sl}{\mathfrak{sl}}
\newcommand{\su}{\mathfrak{su}}
\newcommand{\SU}{\mathrm{SU}}
\newcommand{\SL}{\mathrm{SL}}
\newcommand{\GL}{\mathrm{GL}}
\newcommand{\SO}{\mathrm{SO}}
\newcommand{\Inv}{\mathrm{Inv}}
\newcommand{\mg}{\mathfrak{g}}
\newcommand{\mk}{\mathfrak{k}}
\newcommand{\bB}{\overline{B}}
\newcommand{\tW}{\widetilde{W}}

\renewcommand{\tensor}{\otimes}
\newcommand{\ket}[1]{|#1\rangle}
\newcommand{\bra}[1]{\langle #1|}
\newcommand{\ceil}[1]{\lceil #1 \rceil}
\newcommand{\floor}[1]{\lfloor #1 \rfloor}
\newcommand{\braket}[1]{\langle #1\rangle}
\newcommand{\defeq}{\stackrel{\mathrm{def}}{=}}

\newcommand{\ie}{\textit{i.e.}}
\newcommand{\Ie}{\textit{I.e.}}
\newcommand{\eq}[2]{\begin{equation}\label{#1}#2\end{equation}}

\newcommand{\thm}[1]{Theorem~\ref{#1}}
\newcommand{\cor}[1]{Corollary~\ref{#1}}
\renewcommand{\sec}[1]{Section~\ref{#1}}
\newcommand{\fig}[1]{Figure~\ref{#1}}
\newcommand{\prop}[1]{Proposition~\ref{#1}}
\newcommand{\lem}[1]{Lemma~\ref{#1}}

\newenvironment{fullfigure}[2]
    {\begin{figure}[htb]\begin{center}\def\fullfiga{#1}\def\fullfigb{#2}}
    {\vspace{\baselineskip}\caption{\fullfigb.}\label{\fullfiga}
    \end{center}\end{figure}}

\colorlet{darkblue}{blue!70!black}
\colorlet{darkgreen}{green!70!black}
\colorlet{darkred}{red!70!black}
\colorlet{lightgray}{white!75!black}

\tikzstyle{tangle}=[draw=white,double=darkblue,line width=.08cm,
    double distance=.6pt]

\tikzstyle{stangle}=[style=tangle,scale=.35]
    
\tikzstyle{basic}=[draw=darkblue,fill=darkblue,semithick,scale=.5,>=stealth]

\begin{document}

\title{How hard is it to approximate the Jones polynomial?}

\author{Greg Kuperberg}
\email{greg@math.ucdavis.edu}
\thanks{Partly supported by NSF grants DMS-0606795 and CCF-1013079}
\affiliation{Department of Mathematics, University of
    California, Davis, CA 95616}

\begin{abstract}
\centerline{\textit{\normalsize Dedicated to the memory of
    Fran\c{c}ois Jaeger (1947-1997)}}
\vspace{\baselineskip}
Freedman, Kitaev, and Wang \cite{FKW:simulation}, and later Aharonov,
Jones, and Landau \cite{AJL:approx}, established a quantum algorithm to
``additively" approximate the Jones polynomial $V(L,t)$ at any principal root of
unity $t$.  The strength of this additive approximation depends exponentially
on the bridge number of the link presentation.  Freedman, Larsen, and Wang
\cite{FLW:two} established that the approximation is universal for quantum
computation at a non-lattice, principal root of unity.

We show that any value-distinguishing approximation of the Jones polynomial
at these non-lattice roots of unity is $\shP$-hard.  Given the power to
decide whether $|V(L,t)| < a$ or $|V(L,t)| > b$ for fixed constants $0 <
a < b$, there is a polynomial-time algorithm to exactly count the solutions
to arbitrary combinatorial equations.   Our result is a mutual corollary
of the universality of the Jones polynomial, and Aaronson's theorem that
$\PostBQP = \PP$ \cite{Aaronson:post}.

Using similar methods, we find a range of values $T(G,x,y)$ of the
Tutte polynomial such that for any $c > 1$, $T(G,x,y)$ is $\shP$-hard to
approximate within a factor of $c$ even for planar graphs $G$.

Along the way, we clarify and generalize both Aaronson's theorem and
the Solovay-Kitaev theorem.
\end{abstract}
\maketitle

\section{Introduction}
\label{s:intro}

\nocite{Rowell:paradigm}

A well-known paper of Aharonov, Jones, and Landau \cite{AJL:approx}
establishes a polynomial quantum algorithm to approximate the Jones
polynomial at any principal root of unity; a more abstract form of this
algorithm appeared previously in a paper of Freedman, Kitaev, and Wang
\cite{FKW:simulation}.

\begin{theorem}[Freedman, Kitaev, Wang \cite{FKW:simulation};
    Aharonov, Jones, Landau \cite{AJL:approx}]
Let $t = \exp(2\pi i/r)$ be a principal root of unity, let $L$ be a link
presented by a plat diagram with bridge number $g$, and let $V(L,t)$ be
its Jones polynomial.  Then there is a polynomial-time quantum decision
algorithm that answers yes with probability
$$P[\mathrm{yes}] = \left|\frac{V(L,t)}{(t^{1/2}+t^{-1/2})^{g-1}}\right|^2.$$
\label{th:ajl}
\end{theorem}

(See Burde and Zieschang \cite[\S2.D]{BZ:knots} or \sec{s:braids} for the
definition of a plat diagram and its bridge number.)

In the version of the result of Aharonov et al, the algorithm is jointly
polynomial time in the $r$, the order of the root of unity; as well as in
the bridge number and the crossing number.  They also refine the algorithm
to estimate $V(L,t)$ as a complex number rather than just estimating
its length.  Aharonov et al describe the error in this algorithm as
additive, and note that it would be much harder to provide an algorithm
with multiplicative error. Multiplicative approximation (in the
sense of the complexity class $\APX$ \cite{zoo}) would mean that $V(L,t)$
or $|V(L,t)|$ can be approximated to within some constant factor $c > 1$.

Another way to distinguish between types of error is to say that the
approximation in \thm{th:ajl} is \emph{input-dependent}. Given different plat
diagrams of the same link $L$, the error grows exponentially in one of the
parameters of the presentation, namely the bridge number.  (This additive,
input-dependent model of approximating the Jones polynomial was first
considered in the converse problem of simulating a quantum computer with
the Jones polynomial \cite{BFLW:approx}.)  An algorithm to approximate the
Jones polynomial is only directly useful for topology if the approximation
is \emph{value-distinguishing}; \ie, if there is an error bound which is
independent of quantities other than the value of $|V(L,t)|$.  Multiplicative
approximation is one type of value-distinguishing approximation, but it is
not the most general kind.  For instance, a multiplicative approximation of
$\log(1+|V(L,t)|)$ is much weaker than a multiplicative approximation of
$|V(L,t)|$ itself, but it is still a value-distinguishing approximation.
In general, if an algorithm yield any value-distinguishing approximation
of a real-valued function $f(x)$, it means that for each $c \in \R$, there
exist real numbers $a < c < b$ such that $f(x) < a$ can be distinguished
from $f(x) > b$.  (See also \sec{s:approx}.)

Freedman, Larsen, and Wang \cite{FLW:two} established that the approximated
quantity $|V(L,t)/(t^{1/2}+t^{-1/2})^{g-1}|^2$ in \thm{th:ajl} is
universal for quantum computation when $r=5$ or $r \ge 7$.  Aharonov and
Arad \cite{AA:hardness} establish an $r$-uniform version of this result.
The exceptions, among principal roots of unity, are $t = \exp(2\pi i/r)$ with
$r \in \{1,2,3,4,6\}$.  We call these \emph{lattice} roots of unity, because
they are the roots of unity for which the ring $\Z[t]$ is a discrete subset
of $\C$; the other values of $r$ are \emph{non-lattice} roots of unity.
These results show that even if the approximation is input-dependent,
it is computationally valuable for carefully chosen link diagrams.

On the discouraging side, Vertigan \cite{Vertigan:planar} showed that
it is $\shP$-hard to exactly compute the Jones polynomial $V(L,t)$
except when $t$ is a lattice root of unity.  Jaeger, Vertigan, and Welsh
\cite{JVW:complexity} established a reduction from the Tutte polynomial
of a planar graph to the Jones polynomial of an associated link.  Vertigan
then showed that the specific values of the Tutte polynomial used in this
reduction are $\shP$-hard.

The main result of this article is that the ``encouraging''
universality result strengthens the ``discouraging'' hardness result: Any
value-distinguishing approximation of a value of the Jones polynomial at
a non-lattice root of unity is $\shP$-hard.  The argument is a mash-up
of three standard theorems in quantum computation: The Solovay-Kitaev
theorem \cite{NC:book}, the FLW density theorem, and Aaronson's theorem
that $\PostBQP = \PP$ \cite{Aaronson:post}.  (See also \cite{BFLW:approx}
for a different hardness result.)

\begin{theorem} Let $V(L,t)$ be the Jones polynomial of a link $L$ described
by a link diagram, and let $t$ be a principal, non-lattice root of unity.
Let $0 < a < b$ be two positive real numbers, and assume as a promise that
either $|V(L,t)| < a$ or $|V(L,t)| > b$.  Then it is $\shP$-hard, in the
sense of Cook-Turing reduction, to decide which inequality holds. Moreover,
it is still $\shP$-hard when $L$ is a knot.
\label{th:main} \end{theorem}

\thm{th:main} is proven in \sec{s:pmain} after developing several lemmas.
The theorem is stated for the Jones polynomial and only for values where the
associated braid group representations are unitary and dense.  But the idea
applies to many other link invariants and to many non-unitary values of the
Jones polynomial.  The idea also applies to various functions on graphs or
other input data that aren't link invariants.  We have no formal statement
of a general result, but the basic argument is that if a numerical function
can model the execution of a quantum computer sufficiently accurately,
then typically multiplicative or value-distinguishing approximation is
universal for $\PostBQP$ and therefore $\shP$-hard.  Here is an example
result of this type.

\begin{theorem} Let $c > 1$ and let $x$ and $y$ be two real numbers
such that $q = (x-1)(y-1) > 4$ and $x, y < 0$, and $x$ and $y$ each
have an $\FPTEAS$ approximation.  Then it is $\shP$-hard to 
approximate the Tutte polynomial value 
$T(G,x,y)$ for planar graphs $G$ to within a factor of $c$.
\label{th:second}
\end{theorem}

Here, a real or complex number has an $\FPTEAS$ (fully polynomial-time
exponential approximation scheme) if its digits can be computed in $\FP$,
for instance if it is an algebraic number (\sec{s:approx}).  One interesting
ingredient is that we need the Solovay-Kitaev theorem for non-compact Lie
groups, \thm{th:sk}.  (Aharonov, Arad, Eban, and Landau \cite{AAEL:tutte}
obtained this result for the Lie groups $\SL(d,\R)$ and $\SL(d,\C)$,
which is actually enough for \thm{th:second}.)

We will complete prove \thm{th:second} in \sec{s:psecond}, again
after developing some lemmas.

In related results, Aharonov, Arad, Eban, and Landau \cite{AAEL:tutte}
obtained BQP-universality results about additive approximation to the Tutte
polynomial for planar graphs that are clearly related to \thm{th:second}.
In particular, as with us, their approach involves a study of non-unitary
linear gates.  However, their derivation concerns multivariate Tutte
polynomials, in which different edges of a graph are allowed different
parameters.  The value of $q$ must be the same everywhere, but in their
version the choice of $x$ (say) is taken from a finite list that satisfies
technical conditions.  Following Goldberg and Jerrum, we restrict to a
single pair of values $(x,y)$.

Goldberg and Jerrum \cite{GJ:tutte} showed that multiplicative approximation
of many values of the Tutte polynomial $T(G,x,y)$ is $\NP$-hard (where
the reductions are in $\RP$) for non-planar graphs, while some values
(those with $q=4$ and $-1 < y < 0$) are $\shP$-hard.  Jaeger, Vertigan, and
Welsh \cite{JVW:complexity} also analyzed when $T(G,x,y)$ is $\shP$-hard
to compute exactly.  They noted that the Jones polynomial $V(L,t)$ of an
alternating link $L$ is equivalent to $T(G,x,y)$ for a planar graph $G$
along the curve $xy=1$.  More recently \cite{GJ:planar}, Goldberg and
Jerrum also established that many values of the planar Tutte polynomial
are $\NP$-hard to approximate.  Their new theorems apply to those values
of $(x,y)$ in \thm{th:second} with $q > 5$ (and some other values that we
do not analyze), but their constructions are very different.  Moreover, we
establish $\shP$-hardness, while their planar constructions only establish
$\NP$-hardness.  On the other hand, we use Goldberg and Jerrum's gadget
idea to change from one value of $(x,y)$ to another for a fixed value of $q$.

\begin{remark} The first version of this article contained a significant
mistake, which the reader may grasp after reading \sec{s:postsel}.
The author supposed that all of the implementations of quantum gates
could have complexity $\poly(1/\eps)$ (or $\FPTAS$ approximability) in the
proof of both \thm{th:main} and \thm{th:second}, because this complexity
is sufficient to express the complexity class $\BQP$.  We actually need
complexity $\poly(-\log(\eps))$ (or $\FPTEAS$) to express the complexity
class $\PostBQP$, because this class unavoidably needs exponentially small
probabilities.  Fortunately, the Solovay-Kitaev theorem (\thm{th:sk})
satifies this stringent approximation requirement.   See also \lem{l:ally}
and \thm{th:gateindep} for our corrected constructions.
\end{remark}

\acknowledgments

The author would like to thank Scott Aaronson, Dorit Aharonov, Leslie Ann
Goldberg, and Eric Rowell for helpful discussions.  The author would also
like to thank the referees for their meticulous remarks.

\section{Complexity theory}
\label{s:complexity}

\subsection{Complexity classes}
\label{s:classes}

We assume that the reader is somewhat familiar with complexity classes
such as $\P$, $\NP$, $\BQP$, $\shP$, and the notation that $A^B$ means the
class $A$ with oracle $B$.  See the Complexity Zoo \cite{zoo} and Nielsen
and Chuang \cite{NC:book} for a review.

Whereas a problem in the class $\shP$ counts the number of witnesses
accepted by a verifier in polynomial time, and a problem the class $\NP$
reports whether there is an accepted witness, a problem in the class $\PP$
reports whether a majority of the witnesses are accepted.

\begin{proposition} A problem is $\shP$-hard if and only if it is $\PP$-hard
with respect to Cook-Turing reduction, \ie,
$$\P^\PP = \P^\shP.$$
\label{p:pp} \end{proposition}

\prop{p:pp} is given as an exercise for the reader in the Complexity
Zoo \cite{zoo}.  (Hint: Binary search.)  It is one reason that we use
Cook-Turing reduction in the statement of \thm{th:main}.

A problem which is $\shP$-hard is also hard for the polynomial hierarchy
$\PH$, by the deeper theorem due to Toda \cite{Toda:power} that
$$\PH \defeq \bigcup_{n=1}^\infty
    \underbrace{\NP^{\NP^{\iddots^{\NP}}}}_n \subseteq \P^\shP .$$
The class $\NP$ with a tower of $n-1$ $\NP$s as an oracle is called the
$n$th level of the polynomial hierarchy.  One of the standard conjectures
in complexity theory is the polynomial hierarchy does not collapse, \ie,
that $n$th level does not equal the $n+1$st level for any $n$.  Thus by
Toda's theorem, if a problem is $\shP$-hard, then it is viewed as
qualitatively harder than if it is merely $\NP$-hard.

\subsection{Approximation classes}
\label{s:approx}

The approximation classes listed in the Complexity Zoo \cite{zoo} that
express multiplicative approximation include $\APX$, $\PTAS$, and $\FPTAS$.
These classes are defined there for optimization problems, but they can
equally well be defined for arbitrary functional problems.  Let $f:\Sigma^*
\to \R_+$ be a function that takes bit strings $x$ to positive real numbers.
Then $f(x)$ is in $\APX$ if it can be approximated to within some bounded
factor in polynomial time (with fixed-point output); it is in $\PTAS$
if it can be approximated to within a factor $1+\eps$ in polynomial time
for any $\eps > 0$; and it is in $\FPTAS$ if the computation is jointly
polynomial time in the bit length $|x|$ and $1/\eps$.  (These classes all
refer to deterministic computation; there are analogous randomized classes
such as $\FPRAS$.)

We will need a stricter version of $\FPTAS$.  For many approximate
numerical algorithms, although not usually for optimization problems,
the computation time is jointly polynomial in $|x|$ and $-\log(\eps)$.
We call such an approximation scheme an $\FPTEAS$, or \emph{fully polynomial
time, exponential approximation scheme}.  In particular every algebraic
number has an $\FPTEAS$, using standard numerical algorithms to find
its digits.

Indeed, much more is true:  The digits of algebraic numbers, and the
values of many other elementary functions such as exponentials and
logarithms, can be computed in quasilinear time in the RAM machine model
\cite{Brent:elementary}.  Most numbers that arise in calculus derivations
have quasilinear digit complexity; nearly all of them have polynomial
digit complexity.

We do not know of a standard complexity class to express general
value-distinguishing approximation, so we define such a class here, $\APV$.
Again let $f:\Sigma^* \to \R_+$.  Then $f$ is in $\APV$ if for every constant
$a > 0$, there exists a constant $b > a$ and a polynomial-time algorithm
to decide whether $f(x) > b$ or $f(x) < a$, given the promise that one of
the two is true.  Similarly, we could define a randomized version $\ARV$.
Also, both $\APV$ and $\ARV$ have a variation in which the constant $a$
is an input to one universal algorithm, instead of asking for an algorithm
for each value of $a$.

The following proposition says that if $f(x)$ can be suitably rescaled,
then general value-distinguishing approximation becomes equivalent to
multiplicative approximation in the sense of $\APX$.  \prop{p:apvapx} and
its proof are similar to that of \prop{p:promise}, in particular similar
to the rescaling of Aaronson \cite[Thm 3.4]{Aaronson:post}.  We will need
the contrapositive of \prop{p:apvapx} in the proof of \thm{th:main}.

\begin{proposition} Suppose that $f(x)$ takes positive real values and is in
$\APV$, and suppose further that $|\log(f(x))|$ is bounded by a polynomial in
the bit length $|x|$ of the input.  Suppose that there are constants $c>1$
and $k>1$ such that for every integer $n$, there is a reduction $y_n(x)$
such that
$$f(x) < k^n f(y_n(x)) < c f(x),$$
and suppose that this reduction can be computed in joint polynomial
time in $n$ and in $|x|$.  Then $f(x)$ is in $\APX$.
\label{p:apvapx} \end{proposition}

\begin{proof} Let $a$ and $b$ be some constants such that we can decide by
a subroutine whether $f(x) < a$ or $f(x) > b$ in polynomial time.  Then we
can bound $f(x)$ to within a factor of $cb/a$.  We know by hypothesis that
$f(x) > k^{-m}$ and $f(x) < k^m$ for some $m$ which is polynomial in $|x|$.
So the strategy is to ask whether $f(y_n(x))$ is less than $a$ or more than
$b$ for every $|n| \le m$.  The largest $n$ for which the subroutine reports
that $f(y_n(x)) < a$ yields a good estimate of $ak^{-n}$.  The estimate is
within a factor of $cb/a$, even though the subroutine could give a false
yes answer when $f(y_n(x)) > b$.
\end{proof}

\subsection{Quantum computation}
\label{s:qc}

We cannot give a full review of quantum computation in this article.
There are many equivalent models of quantum computation, and we would
simply like to carefully describe the one that we will use.  Let $D:\Sigma^*
\to \{\yes,\no\}$ be a decision problem, a function $D(x)$ on bit strings
$x$ that takes the values ``yes" and ``no".  In the most 
standard definition of $\BQP$, we assume a uniform family of quantum
circuits $C$ such that $x$ is supplied in input qubits along with
ancillas, and one of the output qubits is the output $D(x)$ with
good probability.   We will use a variation of this definition 
in which the input is encoded in
the circuit rather than in the input to its gates; and
the inputs and outputs are all set to 0.

\begin{proposition} $D \in \BQP$ if and only if there is a quantum circuit
$C = C(x)$ with $\poly(|x|)$ unitary gates acting on $n = \poly(|x|)$
qubits, such that $C$ itself can be generated in deterministic polynomial
time $\FP$, and such that the probability
\eq{e:circuit}{p(x) = |\braket{0^n|C|0^n}|^2}
is at least $\frac23$ if $D(x) = \yes$ and at most $\frac13$ if $D(x)$ is
$\no$.
\label{p:circuit} \end{proposition}

\prop{p:circuit} is a well-known result even though it is not the
most standard definition.  The proof uses the ``uncomputation" method.

\begin{proof} We first assume a circuit $C = C(n)$ of the more standard
type in which $\ket{x}$ is the input along with $\ket{0}$ ancillas, and
one of the qubits is the output.  Then we can make a new circuit $C'$
whose input is all ancillas, and that first changes some of the ancillas
to $\ket{x}$.   One of the outputs $\ket{y}$ of $C'$ agrees
with $D(x)$ with probability at least $\frac23$; the other outputs 
are unpredictable.   We make a new circuit $C''$ that applies $C'$,
then copies $\ket{y}$ to a fresh ancilla with a $\mathrm{CNOT}$ gate,
and then applies $(C')^{-1}$.
\end{proof}

\subsection{Solovay-Kitaev}
\label{s:sk}

In this section we will analyze a central result in quantum computation,
the Solovay-Kitaev theorem.  Let $\BQP_\Gamma$ be the class $\BQP$
defined by some universal finite gate set $\Gamma$.  If each gate in
$\Gamma$ has at least an $\FPTAS$, then the Solovay-Kitaev theorem implies
that $\BQP_\Gamma$ does not depend on the choice of $\Gamma$ and can be
called $\BQP$.  We need some approximability condition here:  If the matrix
entries of gates in $\Gamma$ have intractable or uncomputable information,
then $\BQP_\Gamma$ also carries intractable or uncomputable information
\cite[Thm. 5.1]{ADH:quantum}.

In this paper we will need the more delicate class $\PostBQP$.  As stated
in \thm{th:gateindep}, in order to know that $\PostBQP_\Gamma$ is
independent of $\Gamma$, we need to assume that every gate in $\Gamma$
has an $\FPTEAS$, and not just an $\FPTAS$.  One special case which is
widely used in quantum computation and which we need for \thm{th:main}
is gates with algebraic entries; happily, all algebraic gates have an
$\FPTEAS$.  (Indeed the $\FPTEAS$ class is far more general, as explained
in \sec{s:approx}.)  We also need the Solovay-Kitaev theorem to have
polylogarithmic overhead; happily it does.

Finally, for \thm{th:second} we will need the Solovay-Kitaev theorem for
non-compact Lie groups.  The theorem was originally proven in the case $G
= \SU(d)$.  This case is explained in Nielsen and Chuang \cite{NC:book};
as far as we know the proof works without change when $G$ is any compact,
semisimple Lie group.  Aharonov, Arad, Eban, and Landau \cite{AAEL:tutte}
derive a version of this theorem for the Lie groups $\SL(d,\R)$ and
$\SL(d,\C)$, which are not compact but still semisimple.  Their result is
enough for \thm{th:second}; here we show that the traditional 
argument applies to a more general class of Lie groups.

\begin{theorem}[Solovay, Kitaev] Let $G$ be a connected Lie group whose
Lie algebra $\mg$ is perfect.  Let $\Gamma$ be a finite set of elements
(closed under taking inverses) that densely generates $G$, and let $g
\in G$.  Suppose that there is an $\FPTEAS$ for $g$ and every element of
$\Gamma$. Then there is a word made from $\Gamma$ that approximates $g$,
$$d(g_1g_2\ldots g_m,g) \le \eps,$$
where the length $m$ and the (deterministic) computation time to find the
word are both $\poly(-\log(\eps))$ (non-uniformly in the choice of $G$,
$\Gamma$, and $g$).
\label{th:sk} \end{theorem}

Before turning to the proof of \thm{th:sk}, we discuss some basics of
Lie theory.   (See Varadarajan \cite{Varadarajan:gtm}.)

A Lie group $G$ is a real analytic manifold with a real analytic group law.
(Or a smooth manifold or even just a topological manifold; it turns out
that the group law induces a unique real analytic structure.)  Its Lie
algebra $\mg = T_1G$ is by definition the tangent space at the identity.
We assume that our Lie group $G$ is given with some tractable algorithm
for computing the group law in real analytic coordinates.   For example,
$G$ could be a real algebraic group, by definition a Lie group that can
be realized (non-uniquely) by polynomial equations in some $\GL(n,\R)$.

We can give $G$ a metric to discuss approximation to points in
$G$.  The most natural choice is a left-invariant Riemannian metric
\cite{Paradan:symmetric}.  Every left-invariant Riemannian metric comes
from a positive definite inner product on the Lie algebra $\mg$ of $G$.
Two different inner products on $\mg$ plainly yield different Riemannian
metrics on $G$, but they are they are bi-Lipschitz equivalent.   (If $d_1$
and $d_2$ are two metrics on a set, then they are \emph{bi-Lipschitz
equivalent} if $d_1(p,q) = \Theta(d_2(p,q))$.)  A left-invariant metric on
$\GL(n,\R)$ is not bi-Lipschitz equivalent with Euclidean distance between
matrices, but it is equivalent on any bounded set.  Thus, any of these
choices of metric are equivalent for the purpose of stating \thm{th:sk}.

The usual way to understand the structure of a Lie group $G$ is to begin
with its Lie algebra $\mg$.   A finite-dimensional Lie algebra $\mg$ is
\emph{semisimple} if it is a direct sum of non-abelian, simple Lie algebras.
It is \emph{perfect} if $\mg = [\mg,\mg]$, \ie, $\mg$ is the linear span
of all Lie brackets of pairs of its elements.  (A semisimple Lie algebra
is analogous to a direct product of non-abelian, finite simple groups;
a perfect Lie algebra is analogous to a finite perfect group.)  The most
commonly used Lie algebras, such as $\su(d)$ and $\sl(n,\R)$, have simple
and therefore semisimiple Lie algebras (and are themselves called semisimple
groups).  Every semisimple Lie algebra is perfect, but there are perfect
Lie algebras that are not semisimple.  For example, if $V$ is a linear
representation of a semisimple Lie group $G$ without any trivial summand,
then the Lie algebra of the semidirect product $G \ltimes V$ is perfect.

Every Lie group $G$ has a (real analytic) \emph{exponential map}
$$\exp:\mg \to G$$
defined in polar coordinates by the derivative equation
$$\frac{d}{dt}\exp(tx) = x\exp(tx)$$
for $t \in \R_{\ge 0}$ and $x \in \mg$.  In the special case of an algebraic
group, it is the usual matrix exponential.   We will use three standard
results about the derivative map.  To state the results, we assume some
inner product on $\mg$, and the induced left-invariant metric on $G$.

\begin{proposition} \cite[Thm. 2.10.1]{Varadarajan:gtm} The exponential
map $\exp$ is a bi-Lipschitz, diffeomorphic embedding when
restricted to a ball $B = B(0,\eps)$ of some radius $\eps$ in $\mg$.
\label{p:exp} \end{proposition}

\begin{proposition} \cite[Thm. 2.10.1]{Varadarajan:gtm} Suppose that $\mg$
has a basis $b_1,\ldots,b_k$, and define a function $h:\mg \to G$ by
$$h\left(\sum_j t_j b_j\right) = \prod_j \exp(t_j b_j).$$
Then $f$ is a bi-Lipschitz embedding when restricted to a ball
$B = B(0,\eps)$ of some radius $\eps$ in $\mg$.   Moreover, we 
can choose $\eps$ and $\delta$ so that $f$ is uniformly bi-Lipschitz
for any basis $b'_1,\ldots,b'_k$ with $||b'_j - b_j|| < \delta$.
\label{p:euler} \end{proposition}

\prop{p:euler} is less standard than \prop{p:exp}, but happily Varadarajan
proves a mutual generalization in a single theorem.   The last statement
about uniform constants if the basis $\{b_j\}$ is perturbed is not in the
statement of the theorem, but it follows readily from the proof.
Remark:  The formula in \prop{p:euler} is a generalization of
Euler angles for the group $\SO(3)$.

\begin{proposition} \cite[Thm. 2.12.4]{Varadarajan:gtm} If $[g,h]_G
= ghg^{-1}h^{-1}$ is the group commutator and $[x,y]_\mg$ is the Lie
bracket, then
$$[\exp(x),\exp(y)]_G = \exp([x,y]_\mg + O(\max(||x||,||y||)^3)).$$
\label{p:gbracket}\end{proposition}

Varadarajan proves \prop{p:gbracket} with a less uniform error estimate,
but the same proof establishes the given formula.

The plan of our proof of \thm{th:sk} is not very different from the
standard proof in Nielsen and Chuang \cite{NC:book}:  For some constant
$r < 1$, we create a set of elements in $G$ that, under the inverse of the
exponential map, is a basis of $\mg$ at the scale $r^n$.  In fact, it always
approximately the same basis.  These bases are formed from commutators
at larger scales.   Finally, every element $g \in G$ can first be brought
within the unit ball of the identity and then whittled away to smaller and
smaller scales with these bases.  Since the result is not required to be
uniform in $g$, we do not need a global epsilon net of the Lie group $G$,
only a local one near the identity; a global epsilon net would add extra
difficulties in the non-compact case.   Another trick that simplifies the
derivation is to save the choice of $r$ for the end; it also serves
as a fudge factor to enable the construction.

\begin{proof}[Proof of \thm{th:sk}] Let $k$ be the dimension of $G$.
If $\mg$ is a perfect Lie algebra, then it has a basis $b_1,\dots,b_k$
and elements $x_1,\dots,x_k$ and $y_1,\dots,y_k$ such that $[x_j,y_j]
= b_j$.  We choose some positive definite inner product on $\mg$ and take
the induced left-invariant Riemannian metric on $G$.

By \prop{p:exp}, the exponential map $\exp:\mg \to G$ is a bi-Lipschitz
diffeomorphism within some radius $\eps_1$.   Also, let $\eps_2$ and
$\delta$ be the constants produced by \prop{p:euler}, a radius out to
which the map $f$ is a bi-Lipschitz diffeomorphism.   Also, since the
Lie bracket is bilinear, and by the approximation in \prop{p:gbracket},
we can choose a radius $\eps_3$ within which both the Lie bracket on $\mg$
and the group commutator on $G$ take the ball $B_3 = B(0,\eps_3)$ to itself.
In other words, both brackets are maps
$$[\cdot,\cdot]_\mg:B_3 \times B_3 \to B_3 \qquad
    [\cdot,\cdot]_G:\exp(B_3) \times \exp(B_3) \to \exp(B_3)$$
when $\eps_3$ is small enough.  Finally we choose
$$\eps_0 = \min(\eps_1,\eps_2,\eps_3)$$
to obtain all three properties simultaneously, and we let
$B_0 = B(0,\eps_0)$.

We take advantage of a subtlety of \prop{p:euler}, that the map $h$
only depends on the lines spanned by $\{b_j\}$.  We can thus rescale the
vectors $\{x_j,y_j,b_j\}$ so that they all lie in $B_0$, without disturbing
the constants used to define $B_0$.

We can interpret the group commutator $[\cdot,\cdot]_G$ as a map from $B_0
\times B_0$ to $B_0$ via the equation
$$[x,y]_G \defeq \log([\exp(x),\exp(y)]_G),$$
so that we can then say restate \prop{p:gbracket} as saying that
\eq{e:bracket}{[x,y]_G = [x,y]_\mg + O(\max(||x||,||y||)^3).}

Without loss of generality, $g \in \exp(B_0)$: Because $\Gamma$ densely
generates $G$, we can find a word close to $g$ and multiply $g$ by its
inverse.  Also, we let $r < 1$ be a constant that will be chosen at
the end of the proof.
Again because $\Gamma$ densely generates $G$, we can assume for each $n \le
3$ that it contains the set $\{\exp(b_{j,n})\}$ for a basis $\{b_{j,n}\}$
in $B_0$ such that
\eq{e:berror}{||r^{-n}b_{j,n} - b_j|| < \delta}
for every $j$.   Recall again that $\delta$ is chosen to match \prop{p:euler}.

In the remainder of the proof, we will use asymptotic notation such as
$x = O(r)$ to express errors in Lie elements $x \in \mg$.   What we mean
is that $||x|| < Cr$, where each constant $C$ does not depend on $r$
or $n$, but can depend on everything else defined so far.

For each integer $n \ge 1$, we want to define Lie algebra elements $b_{j,n}$,
$x_{j,n}$, and $y_{j,n}$, all of them words in $\Gamma$ made using the group
law of $G$, such that \eqref{e:berror} holds for all $n$, and such that
\eq{e:xyerror}{x_{j,n}  = r^n(x_j + O(r)) \qquad y_{j,n}  = r^n(y_j + O(r))}
also holds for all $n$.  The definition is by an inductive algorithm
that makes $x_{j,n}$ and $y_{j,n}$ from $b_{j,n+1}$, and makes $b_{j,n}$
from $x_{j,\ceil{n/2}}$ and $y_{j,\floor{n/2}}$.  So the numbering in $n$
is slightly out of order, but since we have already produced $b_{j,n}$
for $n \le 3$, the induction works.

For each $n \ge 1$, we choose integers $t_j = = O(r^{-1})$ so
that the expressions
\eq{e:prod}{\log\left(\prod_j \exp(t_j b_{j,n+1})\right)}
are as close as possible to $r^nx_j$ and $r^ny_j$.  We set $x_{j,n}$
and $y_{j,n}$ to be these approximations.  We claim that the expressions
in \eqref{e:prod} form an $O(r^{n+1}))$-net of $r^nB_0$.   We argue this
in stages:
\begin{description}
\item[1.] The sums $\sum_j t_j r^{n+1}b_j$ are a lattice and an
$O(r^{n+1})$-net by rescaling.
\item[2.] The sums $\sum_j t_j b_{j,n+1}$ are an $O(r^{n+1})$-net because
\eqref{e:berror} limits the distortion of the lattice.
\item[3.] The products $\prod_j \exp(t_j b_{j,n+1})$ are an $O(r^{n+1})$-net
because the map $h$ in \prop{p:euler} is Lipschitz on $B_0$.
\item[4.] The logarithms 
$$\log \left(\prod_j \exp(t_j b_{j,n+1})\right)$$
are an $O(r^{n+1})$-net because the exponential map $\exp$ is inverse
Lipschitz on $B_0$.
\end{description}
Thus, we obtain the error estimates \eqref{e:xyerror}.

For each $n \ge 4$, we let
$$b_{j,n} = [x_{j,\ceil{n/2}},y_{j,\floor{n/2}}]_G.$$
If we combine \eqref{e:xyerror} with \eqref{e:bracket}, we obtain
\eq{e:brerror}{b_{j,n} = r^n(b_j + O(r) + O(r^{3\floor{n/2}-n}))
     = r^n(b_j + O(r)).}
We would like to reconcile \eqref{e:brerror} with \eqref{e:berror}.
The relation \eqref{e:brerror} gives us
$$||r^{-n}b_{j,n} - b_j|| < Cr,$$
and we are done provided that $Cr < \delta$.  So, at final this stage
it is crucial that $C$ does not depend on $n$ or $r$; we can choose
$r$ small enough to make the induction work.

Finally we let $g_0 = g \in \exp(B_0)$.  We inductively let
$$h_n = \prod_j \exp(b_{j,n+1})^{t_j}$$
as in \eqref{e:prod}, and then we let $g_{n+1} = h_n^{-1}g_n$.  We obtain
the estimate
$$||\log(g_{n+1})|| = O(r^{n+1}).$$
It is easy to check by induction that the word length of each $\exp(b_{j,n})$
is $O(n^2)$ (non-uniformly in $r$, but $r$ is now fixed).  Therefore the
word length of the product $h_1h_2\ldots h_n$ is $O(n^3)$.  Also all of
the work to find these words is polynomial in $n$.
\end{proof}

\thm{th:sk} is not uniform in the choice of the group element $g$ and we
do not need this uniformity for our purposes.  However, the proof shows
that it is uniform on any bounded region in $G$.   For completeness,
we give a complementary result that in any semisimple algebraic group,
any element can be efficiently approximated to within a bounded distance.

\begin{theorem} Let $G$ be a semisimple (real) algebraic group which is
equipped with a left-invariant Riemannian metric, and which is densely
generated by a subset $\Gamma$. Let $r > 0$, let $g \in G$, and let $\ell
= d(g,1)$.  Then there is word made from $\Gamma$ that approximates $g$ to within
a bounded distance,
$$d(g_1g_2\ldots g_m,g) < r,$$
with $m = O(\ell+1)$ uniformly in $g$.   Moreover, such a word can
be found in time $\poly(\ell)$.
\label{th:sk2} \end{theorem}

Evidently \thm{th:sk2} can be combined with \thm{th:sk} to obtain a total
word length of
$$m = O(\ell+1)+\poly(-\log(\eps)).$$
Note also that the lower bound $m = \Omega(\ell+1)$ follows from the
triangle inequality
$$d(1,gh) \le d(1,g) + d(1,h)$$
and the fact that the finite set $\Gamma$ has a maximum distance to 1.
So \thm{th:sk2} is optimal up to a constant factor.

We conjecture that \thm{th:sk2} holds for all connected Lie groups.
Note that most named Lie groups, such as $\GL(n,\R)$, $O(n,\C)$, etc.,
are algebraic groups.

\begin{proof} We assume that $G$ is given as a subgroup of some $\GL(n,\R)$
defined by polynomial equations.  We review some of the structure theory
of semisimple real algebraic groups \cite{Paradan:symmetric}:
\begin{description}
\item[1.] $G$ has a maximal compact subgroup $K$.
\item[2.] Every element $g \in G$ has a (canonical) Cartan decomposition
$g = \exp(x)k$, where $k \in K$ and $x \in \mk^\perp \subseteq \mg$.
\item[3.] The quotient manifold $G/K$ has a $G$-invariant Riemannian metric;
it is then called a \emph{symmetric space of noncompact type}.
\item[4.] In the quotient $G/K$, the unique geodesic connecting $gK =
\exp(x)K$ to the identity coset is given by $\exp(tx)K$ with $0 \le t \le 1$.
\item[5.] Up to a change of basis, $G = G^T$, \ie, $G$ is stable under
the transpose map.  $K = G \cap O(n)$ is a maximal compact subgroup if
and only if $G = G^T$.
\item[6.] If $G = G^T$, then the Cartan decomposition $g = \exp(x)k$
coincides with the polar decomposition for matrices, so that $x$ and
$\exp(x)$ are symmetric matrices.
\end{description}
Note also that every $G$-invariant metric on $G/K$ comes from a left-invariant
metric on $G$ which also happens to be right-$K$-invariant.  We assume
such a metric on $G$.  As a consequence, given any two group elements
$g,h \in G$, we have both that
$$d(gK,hK) \le d_G(g,h),$$
and that equality can be achieved by passing to a different representative
$g' \in gK$ or $h' \in hK$.  (We need not change both.)

The idea of our proof is to first find a word with all of the desired
properties in the symmetric space $G/K$ rather than in the group $G$.
The advantage of working in $G/K$ is that we know how to calculate
geodesics and distances, using polar decompositions.   Geometrically,
the idea is not complicated:  We can build a word by taking
steps approximately in the direction of the geodesic from $1K$ to $gK$.

Since $\Gamma$ densely generates $G$, and since closed and bounded
regions in $G$ are compact, we can assume without loss of generality that
$\Gamma$ contains an $r/2$-net of points inside the closed ball $\bB =
\overline{B(1,r)}$ of radius $r$ at the identity.  Given $gK \in G/K$,
let $\gamma$ be the unique geodesic that connects $1K$ to $gK$; we can
compute it from the polar decomposition of $g$.  Let $hK$ be the point at
which $\gamma$ exits $\bB K$.  Then we know or we can assume that
$$d(1K,hK) = d(1,h) = r \qquad d(hK,gK) = d(h,g) = \ell-r.$$
We can choose $g_1 \in \Gamma$ such that $d(g_1,h) < \frac{r}2$.  By the
triangle inequality,
$$d(g_1,g) = d(1,g_1^{-1}g) < \ell-\frac{r}2.$$
Thus, we can let $g' = g_1^{-1}g$ and proceed by induction.

We obtain a word $w$ such that $d(w^{-1}g,K) < \frac{r}2$.  We
are given that $K$ is compact; it follows that there is a finite
set of words $v$ in $\Gamma$ that forms an $r/2$-net of $K$.  
So for one of these words,
$$d(wv,g) = d(v,w^{-1}g) < \frac{r}2 + \frac{r}2 = r,$$
as desired.
\end{proof}

\subsection{Postselection}
\label{s:postsel}

Aaronson \cite{Aaronson:post} defined the class $\PostBQP$ as polynomial-time
quantum computation with free retries, or postselection.  In other words,
the computation can output $\ket{\yes}$, $\ket{\no}$, or $\ket{\retry}$.
(In Aaronson's formal definition, the outputs are measured as $\bra{00}$,
$\bra{01}$, and $\bra{1*}$, respectively; of course the output can equally
well be a qutrit whose values are renamed semantically.)  If the absolute
probabilities are
$$P[\yes] = a \qquad P[\no] = b,$$
then the conditional or postselected probabilities are
$$P[\yes|\mbox{yes or no}] = \frac{a}{a+b} \qquad
P[\no|\mbox{yes or no}] = \frac{b}{a+b}.$$
An algorithm in $\PostBQP$ is required to output ``yes" or ``no" with
conditional (rather than absolute) probability of at least $2/3$.  It is
trivially equivalent to say that for some $c>1$, either $a > cb$ or $b > ca$;
all values of $c$ are equivalent because $c$ can be amplified by repeated
trials.  There is an analogous class $\PostBPP$ for classical randomized
computations; it was also defined previously as $\BPP_{\mathrm{path}}$.
Aaronson established that $\PostBQP = \PP$.  It is not hard to show that
$\PostBQP$ is a subset of $\PP$, just as $\BQP$, $\NP$, and a number of
other important classes are known to be. (The inclusion $\SBQP \subseteq
\AnPP$ is proved in the same way in \prop{p:anpp}.)  The more surprising
fact is that $\PostBQP$ is all of $\PP$.

By contrast, $\PostBPP$ is unlikely to be all of $\PP$.  The relevant
complexity results are as follows:
\begin{description}
\item[1.] $\PostBPP$ contains $\P^{||\NP}$ ($\P$ with parallel $\NP$
queries) \cite{HHT:threshold}.
\item[2.] $\P^{||\NP}$ equals $\P^{\NP[\log]}$ ($\P$ with logarithmically
many $\NP$ queries) \cite{Hemachandra:strong,BH:truth}.
\item[3.] $\PostBPP$ derandomizes to $\P^{||\NP}$. \Ie, they are equal if
sufficiently good pseudo-random number generators exist \cite{SU:approx}.
\item[4.] Without any derandomization assumption \cite{HHT:threshold},
$$\PostBPP \subseteq \BPP^\NP \subseteq \NP^{\NP^\NP}.$$
\end{description}
Thus, $\PostBPP$ is known to be in the third level of $\PH$.  If
we accept derandomization, then it is in the second level.

Another interpretation of $\PostBQP$ or $\PostBPP$ is given
by the following proposition:

\begin{proposition} Let $c > 1$.  Then a decision function $D$ is in
$\PostBPP$ if and only if there are two randomized, polynomial time
algorithms run by Alice and Bob that report ``yes" with probabilities $a$
and $b$, and such that $D(x) = \yes$ when $a > cb$ and $D(x) = \no$ when
$b > ca$.  The same holds for $\PostBQP$ and quantum algorithms.
\label{p:ratio} \end{proposition}

\begin{proof} Suppose that we are given a $\PostBQP$ algorithm in the
original definition.   Then Alice and Bob can both run this algorithm,
with the following conversion:
\begin{align*}
\mbox{yes} &\mapsto \mbox{Alice yes, Bob no} \\
\mbox{no} &\mapsto \mbox{Alice no, Bob yes} \\
\mbox{retry} &\mapsto \mbox{Alice no, Bob no}.
\end{align*}
It is easy to check that this satisfies the requirements of the proposition.
Conversely, suppose that Alice and Bob have separate algorithms.
Then we can combine them into one postselecting algorithm in Aaronson's
sense by flipping a coin to decide which of Alice or Bob runs;
only one of them runs in a given trial.   We can convert according to the
following table:
\begin{align*}
\mbox{Alice yes} &\mapsto \mbox{yes} & \mbox{Alice no} &\mapsto \mbox{retry} \\
\mbox{Bob yes} &\mapsto \mbox{no} & \mbox{Bob no} &\mapsto \mbox{retry}.
\end{align*}
It is easy to check that this conversion satisfies Aaronson's definition.
\end{proof}

We also need to clarify the definition of $\PostBQP$ with regard to different
gate sets.  Aaronson defines $\PostBQP$ using Hadamard and Toffoli gates,
on the argument that all choices of gates are equivalent by Solovay-Kitaev.
But this is somewhat overstated; we give a more precise equivalence
as follows:

\begin{theorem} Let $\Gamma$ be a universal gate set acting on qudits,
let $\PostBQP_\Gamma$ be $\PostBQP$ defined with the gate set $\Gamma$,
and suppose that:
\begin{description}
\item[1.] The matrix entries in each gate have an $\FPTEAS$.

\item[2.] If $z \ne 0$ is expressible as an integer polynomial in the gate
entries with bit complexity $\poly(n)$ with exponents written in unary, then
$$|z| > \exp(-\poly(n)).$$
\end{description}
Then $\PostBQP = \PostBQP_\Gamma$.  If only condition 1
holds, then $\PostBQP \subseteq \PostBQP_\Gamma$.
\label{th:gateindep} \end{theorem}

Before proving \thm{th:gateindep}, here are three remarks.  First, the
class $\BQP$ only requires a weaker version of condition 1, namely that
each gate in $\Gamma$ has an $\FPTAS$, in order to enable the Solovay-Kitaev
theorem.  We need $\FPTEAS$ because $\PostBQP$ relies on exponentially small
probabilities.  Without exponentially good approximation, Solovay-Kitaev
would still give us a circuit reduction, but the reduction would be
relative to $\P/\poly$ rather than relative to $\P$.  Second, we conjecture
that if only condition 1 holds, then $\PostBQP$ and $\PostBQP_\Gamma$
are not always equal.  Third, we do not know whether postselected quantum
computation is gate-independent with a time bound of $\tilde{O}(n^\alpha)$
for some fixed exponent $\alpha$, because the Solovay-Kitaev theorem could
change the exponent.

\begin{proof} Condition 1 and \thm{th:sk} together imply that $\PostBQP
\subseteq \PostBQP_\Gamma$.  The traditional gate set consisting of Hadamard
and Toffoli gates can be approximated using gates in $\Gamma$; how good of
an approximation is sufficient?   It is easy to check that the Hadamard and
Toffoli gates satisfy condition 2, so the strength of approximation that
we need is $\exp(-\poly(|x|))$.  This is precisely how much \thm{th:sk}
gives us with polynomial overhead, if each gate in $\Gamma$ has an $\FPTEAS$.

The same argument works in reverse, but we must add condition 2
explicitly, since it is not guaranteed in general.
\end{proof}

We will not strictly need the following proposition, but it helps
for understanding \thm{th:gateindep}.  It shows that any gate set with
algebraic matrix entries automatically satisfies condition 2.

\begin{theorem} Let $t_1,\ldots,t_k$ be a finite list of algebraic
numbers in $\C$, and let $p$ be an integer polynomial in $k$ variables with
bit complexity $\poly(n)$ with exponents written in unary.   Then
$$|p(t_1,\ldots,t_k)| > \exp(-\poly(n))$$
(non-uniformly in the choice of $\{t_j\}$), assuming that the
value is non-zero.
\label{th:nearmiss} \end{theorem}

\begin{proof} We first reduce to the case $k=1$.  The numbers $\{t_j\}$
all lie in some finite-degree field extension $K \supseteq \Q$.   It is
a theorem of Galois that every such field has a generator $t$.   We thus
obtain that each $t_j = p_j(t)$ is some rational polynomial in $t$, and by
rescaling $t$, we can make each $p_j$ an integer polynomial; these fixed
polynomials can be composed with the polynomial $p$ in the proposition.
Thus, without loss of generality, we can take $k=1$ and $t = t_1$.

Next we consider the case that $t = \frac{a}{b} \in \Q$ is rational.
In this it is enough for $p$ to have degree $\poly(n)$, because
we immediately get
$$|p(t)| > b^{\deg p}.$$
In the general case, let $d$ be the degree of the field $K$, and
let $z = p(t)$.   Then $z = z_1$ has a list of Galois conjugates
$z_1,z_2,\ldots,z_d$.  Moreover, if we choose some basis of the
ring of integers of $K$, then $t$ has rational coordinates $s_1,\ldots,s_d$,
and we can write
$$\prod_{j=1}^d z_j = q(s_1,\ldots,s_d)$$
for a polynomial $q$ with $\deg q = d(\deg p)$.  Thus by the rational
case we obtain
$$\biggl|\prod_{j=1}^d z_j\biggr| > \exp(-\poly(n)).$$
At the same time, because of the degree bound on $p$ and because
each coefficient of $p$ is bounded by $\exp(\poly(n))$, we obtain
$$|z_j| < \exp(-\poly(n)).$$
By dividing through, we obtain
$$|z| = |z_1| > \exp(-\poly(n)).$$
\end{proof}

It is important to compare $\PostBQP$ and $\PostBPP$ to three other
complexity classes:  $\AnPP$, or one-sided almost wide $\PP$, defined
by Vyalyi \cite{Vyalyi:qma}; $\SBP$, or small-bounded probabilistic
$\P$ \cite{BGM:error}; and a quantum class that we will call $\SBQP$.
All three classes depend on a real-valued function $f(x)$ in $\FP$
(expressed in fixed-point arithmetic, say), where $x$ is the input to the
decision problem, and a constant $c > 1$.  The classes $\SBP$ and $\SBQP$
are defined in the same way as the Alice-Bob definition of $\PostBPP$ and
$\PostBQP$, except with a different model for Bob.  As in \prop{p:ratio},
Alice executes a randomized algorithm in the case of $\SBP$ and a quantum
algorithm in the case of $\SBQP$ and has success probability $a$.  Meanwhile
Bob's value $b = f(x)$ is computed directly in $\FP$, as a real number in
fixed-point arithmetic.  In both $\SBP$ and $\SBQP$, the answer is ``yes"
when $a > cb$ and ``no" when $b > ca$.

Finally, $\AnPP$ is a non-quantum class that is closely related to $\PP$
and is defined similarly to $\SBP$.  Like $\SBP$, a decision function $D
\in \AnPP$ has a function $b = f(x)$ which lies in $\FP$, and a randomized
algorithm whose success probability is $a$.  When $D \in \AnPP$, we
require that
\begin{align*}
D(x) = \mathrm{yes} &\implies a > cb+\frac12 \\
D(x) = \mathrm{no} &\implies \frac12 \le a < b+\frac12,
\end{align*}
which again is like $\SBP$ but has an extra $\frac12$ term.

\begin{lemma} Without loss of generality, the function $f(x)$ in the
definition of $\AnPP$, $\SBP$, $\SBQP$ can be taken to be $2^{-p(|x|)}$
for some $p$; and all values of the constant $c$ are equivalent.
\label{l:sbqp} \end{lemma}

\begin{proof} The constant $c$ is irrelevant by the usual technique of
amplification by repeated trials.  This is immediate in the case of $\SBP$
and $\SBQP$.  It is not very difficult in the case of $\AnPP$, and was
established by Vyalyi \cite{Vyalyi:qma}.

To argue that $f(x)$ can be set to $2^{-p(|x|)}$ (in the cases of $\SBP$
and $\SBQP$), first choose $p$ so that $f(x) > 2^{-p(|x|)}$.  Then Alice
can compute $f(x)$ and reduce her success probability by a factor of
$2^{p(|x|)}f(x)$.  The argument in the case of $\AnPP$ is essentially the
same and was also explained by Vyalyi \cite{Vyalyi:qma}.
\end{proof}

\begin{proposition}
$$\SBQP = \AnPP.$$
\label{p:anpp} \end{proposition}

\begin{proof} The proof is almost the same as Aaronson's proof that
$\PostBQP = \PP$ \cite[Thm. 3.4]{Aaronson:post}.   We can also define $\AnPP$
as a counting class in which, for each certificate $y$ of length $n$, the
computation produces a value $f(y) = \pm 1$, and these values are summed
to produce $A(x)$.  For a decision problem $D \in \AnPP$, we require that
\begin{align*}
D(x) = \mathrm{yes} &\implies A(x) > 2^nCb \\
D(x) = \mathrm{no} &\implies 0 \le A(x) < 2^nb.
\end{align*}

First, let $L \in \SBQP$ be computed by a quantum circuit that consists of
Hadamard and Toffoli gates. It is convenient to change the counting model
of $\AnPP$ slightly to let the values be $\pm 1$ or $0$.  Then we obtain
an $\AnPP$ algorithm by multilinear expansion of the effect of these gates
on density matrices.  The matrix entries of a Toffoli gate, in its effect
on a density matrix, are $0$ and $1$; the corresponding matrix entries
of a Hadamard gate are $\pm \frac12$.  The final probability is given
by a partial trace of the output density matrix, and is non-negative and
exactly matches the criteria for $\AnPP$.

Now let $L \in \AnPP$ and let $a$ be Alice's success probability in the
$\AnPP$ algorithm.  We can again slightly re-express the counting model
of $\AnPP$ so that $f(y) \in \{0,1\}$ and its sum $A = A(x)$ is given by
$A = 2^na$.

Then, in the $\SBQP$ algorithm, we can quantum-compute the unitary map
$$U_f\ket{y} = \ket{y,f(y)},$$
where the value $f(y)$ is written to an ancilla qubit.  We provide the
input $\ket{++\cdots+}$ to $U_f$, and then postselect on whether the left
$n$ qubits of the result are all $\ket{+}$.  If they are, then the ancilla
qubit has the state
$$\ket{\psi} \propto (1-a)\ket{0} + a\ket{1}.$$
If this qubit is measured in the $\pm$ basis, then the probability
of $\ket{-}$ is
$$a' = \frac{(2a-1)^2}{1+(2a-1)^2}.$$
If we assume that $b > \frac1{4}$ and let $c = 2$ in the $\AnPP$ algorithm,
then
\begin{align*}
0 < a < \frac12 + b &\implies a' < \frac{4b^2}{1+4b^2} < 4b^2 \\
a > \frac12 + 2b &\implies a' > \frac{16b^2}{1+16b^2} > 8b^2.
\end{align*}
So we can let $b' = 4b^2$ and $c' = 2$ in an $\SBQP$ algorithm
that produces the probability $a'$.
\end{proof}

Many of the complexity classes discussed here employ the semantic condition
that the probabilities of particular outcomes are above one threshold or
below another threshold.  We can also consider promise versions of these
classes in which these conditions hold for some inputs and not others.
When they are considered in promise form, $\SBP$- and $\SBQP$-hardness 
are the same as $\PostBPP$- and $\PostBQP$-hardness.  The 
non-trivial part of this equality (given that $\SBP \subseteq \PostBPP$
and $\SBQP \subseteq \PostBQP$) is the following inclusions:

\begin{proposition}
\begin{align*}
\PromisePostBPP &\subseteq \P^{\PromiseSBP} \\
    \PromisePostBQP &\subseteq \P^{\PromiseSBQP}.
\end{align*}
\label{p:promise} \end{proposition}

\begin{proof} Suppose that $D \in \PromisePostBQP$ is a decision function and
that it is implemented by a quantum circuit.  We recall the assumption that
$$\max(a,b) > 2^{-n},$$
where $n = \poly(|x|)$ and $x$ is the input.

The construction is then similar to a rescaling argument in Aaronson's
proof that $\PostBQP = \PP$ (explained in \cite[Thm. 3.4]{Aaronson:post}
in the second half of the main proof).  We assume that either $a > 8b$
or that $b > 8a$.  Then for each $0 \le k \le n$, use $\PromiseSBQP$ to
compare both $a$ and $b$ to $2^{-k}$.  If $a > 8b$, then for every $k$,
$\PromiseSBQP$ will either reliably report that $a > 2^{-k}$ or that $2^{-k}
> b$, and there will exist a $k$ for which it will do both.  Meanwhile if
$b > 8a$, it will report that $b > 2^{-k}$ or that $2^{-k} > a$, and both
for at least one $k$.  These two outcomes are mutually exclusive.

The argument that $\PromisePostBPP \subseteq \P^{\PromiseSBP}$ is the same,
but simpler since the lower bound on $\max(a,b)$ is immediate.
\end{proof}

Finally, as noted by Aaronson, linear computation is another interesting
interpretation of $\PostBQP$.  (This is linear computation in the sense of
non-unitary quantum computation, not $\Z/2$-linear circuits or numerical
linear algebra!)  Post-conditioning allows us to replace unitary gates by
subunitary gates, and to rescale subunitary gates arbitrarily.  But every
linear operator that acts on vector states $\ket{\psi}$ is proportional to
a subunitary operator.  Thus, $\PostBQP$ can also be defined by the class of
polynomial-sized circuits with linear gates, without the unitary restriction.

At first glance, the measurement probability \eqref{e:circuit} used for
$\PostBQP$ still use the Hilbert space structure, even if the gates do not.
But this is not entirely true either.  If circuits are evaluated in a
form such as $\braket{0^n|C|0^n}$, and if the gates need not be unitary,
then there is no need to equate the vector $\ket{0}$ with the dual vector
$\bra{0}$ using a Hermitian form.  We can instead define $\bra{0}$ and
$\bra{1}$ to be the dual basis to $\ket{0}$ and $\ket{1}$.  The drawback to
this computational model is that it does not have a reasonable notion of a
mixed state, nor partial trace that makes mixed states from pure states.
We may define $\bra{0}$ using both $\ket{0}$ and $\ket{1}$ (using the
relations $\braket{0|0} = 1$ and $\braket{0|1} = 0$), but we cannot in
general define $\bra{\psi}$ or $\ket{\psi}\bra{\psi}$ from $\ket{\psi}$.

Indeed, we can more cleanly define linear computation as computation with
\emph{libits} (linear bits).  By definition, a libit is like a qubit in the
sense that it is assigned a 2-dimensional complex state space $V$.  But unlike a
qubit, $V$ is just a vector space with no Hilbert space structure, so
that there isn't even any way to say whether linear gates acting on libits
are unitary.  A libit has kets, which are vectors $\ket{\psi} \in V$,
and it has bras, which are dual vectors $\bra{\psi} \in V^*$.  But $V$
and $V^*$ are simply different vector spaces.

\section{The Jones polynomial}
\label{s:jones}

In this section we review the definition of the Jones polynomial and some
theorems about it that lead to a proof of \thm{th:main}.  We will define
the Jones polynomial using the Kauffman bracket formalism, which in our
opinion is one of the simplest and nicest definitions.  For background
see Kauffman \cite{Kauffman:spinknot}; also previous work by the author
\cite[\S2]{Kuperberg:spiders} has a review of properties of the Kauffman
bracket renamed as the ``$A_1$ spider".

\subsection{The Kauffman bracket}

Let $t^{1/4} \in \C^\times$ be a non-zero complex number.  (The reason
for this notation is that all of the essential mathematics of the Jones
polynomial depends only $t$, even though it is convenient to choose a fourth
root $t^{1/4}$ to define it.)  Then the Kauffman bracket is defined as a
function on links projections, or \emph{link diagrams}, by the following
recursive relations:
\begin{align}
\left|\begin{tikzpicture}[baseline=-.75ex,style=stangle]
\useasboundingbox (-.6,.6) rectangle (.6,.6);
\draw[double] (-.5,.5) -- (.5,-.5);
\draw[double] (-.5,-.5) -- (.5,.5);
\end{tikzpicture}\right\rangle_K
&= -t^{1/4}
\left|\begin{tikzpicture}[baseline=-.75ex,style=basic]
\useasboundingbox (-.6,.6) rectangle (.6,.6);
\draw (-.5,-.5) arc (-45:45:.707);
\draw (.5,.5) arc (135:225:.707);
\end{tikzpicture}\right\rangle_K
-t^{-1/4}
\left|\begin{tikzpicture}[baseline=-.75ex,style=basic]
\useasboundingbox (-.6,.6) rectangle (.6,.6);
\draw (.5,-.5) arc (45:135:.707);
\draw (-.5,.5) arc (225:315:.707);
\end{tikzpicture}\right\rangle_K \label{e:skein} \\
\left\langle\begin{tikzpicture}[baseline=-.75ex,style=basic]
\useasboundingbox (-.6,.6) rectangle (.6,.6);
\draw (0,0) circle (.5);
\end{tikzpicture}\right\rangle_K
&= -t^{1/2}-t^{-1/2}. \nonumber
\end{align}
Relations of this type are called \emph{skein relations}.  (Kauffman writes
\eqref{e:skein} with a bracket $\braket{\cdot}$ for all terms, but a ``ket"
is more consistent with standard quantum notation; see \sec{s:skein}.)
What the relations mean is that if three link diagrams $L_1$, $L_2$, and
$L_3$ are identical except that they differ in one place as indicated,
then their Kauffman bracket values satisfy the given linear relation:
$$\braket{L_1}_K = -t^{1/4} \braket{L_2}_K - t^{-1/4} \braket{L_3}_K.$$
The second equation says that if $L_1$ and $L_2$ are two link diagrams
that are the same except that $L_1$ has an extra circle, then
$$\braket{L_1}_K = -(t^{1/2}+t^{-1/2}) \braket{L_2}_K.$$
The base of the recursion is given by saying that the Kauffman bracket of
the empty link diagram is 1.  With this normalization, the Jones
polynomial is given by
$$V(L,t) = \frac{\braket{L}_K}{-(t^{1/2}+t^{-1/2})t^{3w/4}},$$
where $w$ is the writhe of the diagram $L$, \ie, the number of positive
crossings minus the number of negative crossings.  It is a remarkable fact,
although it is not difficult to check, that the Kauffman bracket is invariant
under the second and third Reidemeister moves \cite[\S1.C]{BZ:knots},
and that the Jones polynomial is invariant under all three Reidemeister
moves and is therefore a link invariant.

\subsection{Skein spaces}
\label{s:skein}

The importance of the skein relations is that they can be extend the
Kauffman bracket to a ``Kauffman ket" for tangles.  Here a \emph{tangle}
is an incomplete link, \ie, the intersection of a link and a ball whose
boundary is transverse to the link.  By definition, the Kauffman ket of
a tangle is a vector in a corresponding \emph{skein space}; actually the
skein space itself is defined from the tangles.  More precisely, given a
3-dimensional ball with $2n$ marked points, let $F(2n)$ be the formal vector
space of linear combinations of all tangles that end at the marked points.
Then the skein space $W(2n) = F(2n)/\sim$ is by definition the quotient
of the vector space $F(2n)$ by the relations \eqref{e:skein}.  Any
element of $W(2n)$, \ie, any linear combination of tangles modulo
the skein relations, is called a \emph{skein}.  In this
construction, then, the Kauffman bracket $\ket{T}$ of a tangle $T$
is ``itself", \ie, the skein that it represents.  If $W(2n)$ is
a skein space of tangles with $2n$ endpoints, then the Kauffman relations
imply that
\eq{e:kcat}{\dim W(2n) = C_n = \frac{1}{n+1}\binom{2n}{n},}
the $n$th Catalan number, because the planar matchings of the $2n$ endpoints
are a basis of the skein space.

When the parameter $t$ is a root of unity, it is more important to look at
a certain reduced skein space $X(2n)$.  First, we take an explicit model of
$W(2n)$ as the skein space of tangles in the right half-plane with end points
at the integers $1,2,\ldots,2n$ on the vertical number line.  Then there is
another skein space $W'(2n)$ consisting of tangles in the left half plane
and with the same boundary.  ($W'(2n)$ is of course equivalent to $W(2n)$,
but in more than one way: by reflection, by rotation by 180 degrees, etc.)
Then there is a bilinear pairing
$$\braket{\cdot,\cdot}_K:W(2n) \times W'(2n) \to \C$$
given by gluing together one tangle on each side and evaluating the
Kauffman bracket.  For example:
$$
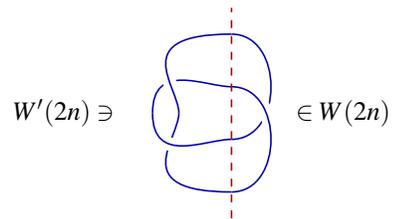
\begin{tikzpicture}[style=stangle]
\draw[double] (-3,0) .. controls (-3,2) and (-1,1) .. (0,1);
\draw[double] (-2,0) .. controls (-2,1) and (-4,3) .. (0,3);
\draw[double] (-2,0) .. controls (-2,-1) and (-4,-3) .. (0,-3);
\draw[double] (-3,0) .. controls (-3,-2) and (-1,-1) .. (0,-1);
\draw[double] (0,3) .. controls (2,3) and (2,-1) .. (0,-1);
\draw[double] (0,-3) .. controls (2,-3) and (2,1) .. (0,1);
\draw[semithick,darkred,dashed] (0,-4) -- (0,4);
\draw (-4,0) node[anchor=east] {$W'(2n) \ni $};
\draw (2,0) node[anchor=west] {$\in W(2n)$};
\end{tikzpicture}.$$
Finally,
$$X(2n) \defeq W(2n)/(\ker \braket{\cdot,\cdot}_K).$$
It is known that $\braket{\cdot,\cdot}_K$ is degenerate on $W(2n)$ if and
only if $t$ is a root of unity of order $r > 1$ and $n \ge r-1$.  Moreover,
if $|t| = 1$, then there is a conjugate-linear isomorphism between $W(2n)$
and $W'(2n)$ given by reflecting the tangle across the horizontal line.
(The reflection reverses crossings, so we need $|t| = 1$ in order to have
$t^* = t^{-1}$ and thus have conjugate linearity.)  Thus, if $|t| = 1$,
then $\braket{\cdot,\cdot}_K$ is a non-degenerate Hermitian form on the
quotient space $X(2n)$.  It is further known that $\braket{\cdot,\cdot}_K$
is positive definite if $t = \exp(2\pi i/r)$ is a principal root of unity.
Thus, if $t$ is a principal root of unity, $X(2n)$ is a finite-dimensional
Hilbert space, so it and the Jones polynomial become relevant to quantum
computation.  (See \sec{s:other} for references and further explanation.)

The skein spaces $W(2n)$ and $X(2n)$ have an action of the braid group
$B_{2n}$ on $2n$ strands.  The action is given by attaching the braid to
a tangle or skein to make a new tangle or skein:
$$
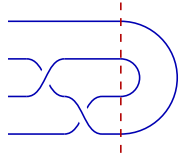
\begin{tikzpicture}[style=basic,baseline=.6cm]
\foreach \x in {0,1,2,3} { \draw (-1,\x) -- (2,\x); }
\draw (2,0) arc (-90:90:1.5);
\draw (2,1) arc (-90:90:.5);
\foreach \x/\y/\s in {-.5/1/1, .5/0/0}
{ 
    \begin{scope}[xshift=\x cm,yshift=\y cm,style=tangle]
    \fill[white] (.02,-.05) rectangle (.98,1.05);
    \draw[double] (0,\s) .. controls (.5,\s) and (.5,1-\s) .. (1,1-\s);
    \draw[double] (0,1-\s) .. controls (.5,1-\s) and (.5,\s) .. (1,\s);
    \end{scope}
}
\draw[dashed,darkred] (2,-.5) -- (2,3.5);
\end{tikzpicture}$$
This is the \emph{Jones braid representation} on $X(2n)$ \cite{FLW:two}.
In key cases $X(2n)$ is a Hilbert space and the braid representation is
unitary (\sec{s:other}).

A variation of this theme is that if $\sigma \in B_n$ is a braid on $n$
strands, we can simply expand it as a skein in $W(2n)$, with $n$ endpoints
on the left and on the right.  (Or in $X(2n)$, but for the moment $W(2n)$ is
more relevant.)  We can also concatenate two elements of $W(2n)$ in the same
way that braids are multiplied.  \Ie, having segregated the $2n$ endpoints
into $n$ each on the left and right, we can define a bilinear product map
$$m:W(2n) \times W(2n) \longto W(2n),$$
where $m(s,t)$ is given by attaching the right endpoints of $s \in W(2n)$
to the left endpoints of $t \in W(2n)$.  This makes $W(2n)$ into an
associative algebra called the Temperley-Lieb algebra \cite{AJL:approx}.
The Jones braid representation generalizes to a representation
$$\rho:W(4n) \times X(2n) \longto X(2n)$$
of the Temperley-Lieb algebra $W(4n)$, given by attaching $s \in W(4n)$
to $t \in X(2n)$ along half of the endpoints of the former and all of the
endpoints of the latter.

\subsection{Other models of skein spaces}
\label{s:other}

There are many ways to define the skein space $W(2n)$ and the reduced
skein space $X(2n)$, and the braid group action on them.  One of the
most important models is that, when $t$ is not a root of unity, $W(2n)$
is the invariant subspace $\Inv(V^{\tensor 2n})$ of the representation
$V^{\tensor 2n}$ of the quantum group $U_{\sqrt{t}}(\sl(2))$, where $V$
is the standard 2-dimensional irreducible representation \cite{Kassel:gtm}.
This model is well-known to be the equivalent to the Kauffman skein space
that we use here \cite{FK:canonical}.  Moreover, it is well-known that as $t$
approaches a principal root of unity, the pairing $\braket{\cdot,\cdot}_K$
on $W(2n)$ undergoes a degeneration, that the reduced skein space $X(2n)$
is a Hilbert space, and that the associated braid representation is
unitary \cite{Kirillov:inner,Wenzl:tensor}.  In fact, all of these facts
are part of a larger theory for all quantum groups $U_{\sqrt{t}}(\mg)$
for any simple Lie algebra $\mg$.  Unfortunately, it is not practical to
give a summarize the theory of quantum groups here.

Since Aharonov, Jones, and Landau \cite{AJL:approx} use the so-called path
model, we want to relate our planar matchings model to that one.  In any
case, the path model helps to compute the dimension of $X(2n)$, and it
yields one proof that it is a Hilbert space.  The rest of this section is
a summary of calculations based on more advanced points of the Kauffman
skein theory \cite{KL:recoupling}.  We do not include complete proofs.
The results are not needed for our results, other than the one standard
fact that $X(2n)$ is a Hilbert space when $t$ is a principal root of unity.

Model $W(2n)$ with planar matchings in the upper half plane.  These are
equivalent to balanced strings of parentheses, by matching the parentheses:
$$
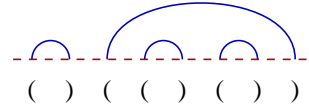
\begin{tikzpicture}[style=basic]
\draw (1,0) arc (0:180:.5);
\draw (7,0) .. controls (7,2) and (2,2) .. (2,0);
\draw (4,0) arc (0:180:.5);
\draw (6,0) arc (0:180:.5);
\draw[dashed,darkred] (-.5,0) -- (7.5,0);
\draw (0,-.3) node[anchor=north] {(};
\draw (1,-.3) node[anchor=north] {)};
\draw (2,-.3) node[anchor=north] {(};
\draw (3,-.3) node[anchor=north] {(};
\draw (4,-.3) node[anchor=north] {)};
\draw (5,-.3) node[anchor=north] {(};
\draw (6,-.3) node[anchor=north] {)};
\draw (7,-.3) node[anchor=north] {)};
\end{tikzpicture}$$
Then, a balanced string of parentheses of length $2n$ is equivalent to a
path from $0$ to $0$ in the non-negative integers $\Z_{\ge 0}$, given by
stepping to the right at each left parenthesis and to the left at each
right parenthesis.

It is known that the planar matchings corresponding to the paths that lie in
the discrete interval $\{0,1,\ldots,r-2\}$ are a basis of $X(2n)$, when $t$
is an $r$th root of unity with $r>1$.  Call these the \emph{admissible}
matchings.  They are not an orthogonal basis, but their Gram-Schmidt
orthogonalization in a natural partial ordering is the path basis used in
\cite{AJL:approx}.  (In other words, the admissible matchings are those
whose parentheses do not nest beyond a depth of $r-2$.)  The partial
ordering can be expressed as a relation on paths, that $p \succeq q$
if the path $p$ never crosses to the right of $q$.

In order to argue these facts, one employs a special skein with $2n$
endpoints called a \emph{Jones-Wenzl projector}, which is given by the
following recurrence relation
$$\begin{tikzpicture}[style=basic,baseline=-1ex]
\draw[black] (-.5,0) node[anchor=south] {\small $n$};
\draw (-1,0) -- (1,0);
\draw[darkred,fill=white] (-.1,-.5) rectangle (.1,.5);
\end{tikzpicture} = 
\begin{tikzpicture}[style=basic,baseline=-1ex]
\draw[black] (-.75,0) node[anchor=south] {\small $n-1$};
\draw (-1,0) -- (1,0);
\draw[darkred,fill=white] (-.1,-.5) rectangle (.1,.5);
\draw (-1,-.7) -- (1,-.7);
\end{tikzpicture} + 
\frac{[n-1]}{[n]}
\begin{tikzpicture}[style=basic,baseline=-1ex]
\draw (0,.3) -- (2,.3);
\draw (-1,0) -- (0,0);
\draw (2,0) -- (3,0);
\draw (-1,-.7) -- (0,-.7) .. controls (1,-.7) and (1,-.3) .. (0,-.3);
\draw (3,-.7) -- (2,-.7) .. controls (1,-.7) and (1,-.3) .. (2,-.3);
\draw[black] (-.75,0) node[anchor=south] {\small $n-1$};
\draw[black] (1,.3) node[anchor=south] {\small $n-2$};
\draw[darkred,fill=white] (-.1,-.5) rectangle (.1,.5);
\draw[darkred,fill=white] (1.9,-.5) rectangle (2.1,.5);

\end{tikzpicture},$$
and the rule that the projector of order 1 is a plain strand.  Here a 
strand labeled with $n$ means $n$ strands, and $[n]$
is a \emph{quantum integer} defined by the formula
$$[n] = \frac{t^{n/2} - t^{-n/2}}{t^{1/2} - t^{-1/2}}.$$
The Jones-Wenzl projector exists for all $n$ when $t$ is not a root of
unity or $t=1$, and it exists when $n < r$ when $t$ is a root of unity of
order $r > 1$.  Also, the projector of order $r-1$ vanishes in $X(2r-2)$.
When working with reduced skein spaces $X(2k)$, we can assume, as a new
skein relation, that the projector of order $r-1$ vanishes.  This new
skein relation allows us to express a planar matching whose path reaches
$r-1$ in terms of earlier planar matchings.  Thus, we can conclude that
the admissible matchings are a spanning set of $X(2n)$, and we can ignore
the inadmissible matchings.

Then, we can modify a planar matching by inserting a vertical projector
between every pair of endpoints:
$$
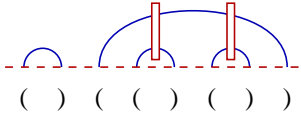
\begin{tikzpicture}[style=basic]
\draw (1,0) arc (0:180:.5);
\draw (7,0) .. controls (7,2) and (2,2) .. (2,0);
\draw (4,0) arc (0:180:.5);
\draw (6,0) arc (0:180:.5);
\draw[dashed,darkred] (-.5,0) -- (7.5,0);
\draw (0,-.3) node[anchor=north] {(};
\draw (1,-.3) node[anchor=north] {)};
\draw (2,-.3) node[anchor=north] {(};
\draw (3,-.3) node[anchor=north] {(};
\draw (4,-.3) node[anchor=north] {)};
\draw (5,-.3) node[anchor=north] {(};
\draw (6,-.3) node[anchor=north] {)};
\draw (7,-.3) node[anchor=north] {)};
\draw[darkred,fill=white] (3.4,.2) rectangle (3.6,1.7);
\draw[darkred,fill=white] (5.4,.2) rectangle (5.6,1.7);
\end{tikzpicture}$$
(The projectors of order 0 and 1 can be omitted, since they are trivial.)
Call a skein of this form a \emph{path vector}.  By expanding the projectors,
one can show that path vectors are related to admissible planar matchings
by a triangular matrix.  Since admissible matchings span $X(2n)$, so do the
path vectors; and if the path vectors are linearly independent in $X(2n)$,
so are admissible matchings.

The path vectors, as vectors in $W(2n)$ and $W'(2n)$, have a Gram matrix
using the bilinear form on these two spaces.  It is not hard to check,
using various properties of Jones-Wenzl projectors, that this Gram matrix
is diagonal and that the diagonal entries are non-zero.  Thus, the path
vectors are a basis of $X(2n)$.  When $t$ is a principal root of unity,
the diagonal entries are also positive real numbers, which implies that
$X(2n)$ is a Hilbert space.  Finally, the triangular change of basis
from admissible matchings to path vectors shows that the latter are the
Gram-Schmidt orthogonalization of the former.

\subsection{Quantum computation with braids}
\label{s:braids}

The idea, first explained by Freedman, Larsen, and Wang \cite{FLW:universal}
is that when $t$ is a principal root of unity, the Hilbert space $X(2n)$
can be interpreted as a quantum memory, and a braid $\sigma \in B_{2n}$ can
be interpreted as a quantum circuit.  The question then is whether such a
model is universal for quantum computation.  The well-known answer is yes
when $t$ is a non-lattice, principal root of unity, and the main technical
tool is the following theorem.

\begin{theorem}[Freedman, Larsen, Wang \cite{FLW:two}] Let $t = \exp(2\pi
i/r)$ with $r=5$ or $r \ge 7$.  Then Jones braid representation of $B_{2n}$
is dense in $\PSU(X(2n))$ for $n \ge 2$, or for $n \ge 3$ in the case $r=10$.
\label{th:flw} \end{theorem}

\begin{corollary} Let $t = \exp(2\pi i/r)$ with $r=5$ or $r \ge 7$. Let
$$p(x) > 2^{-\poly(|x|)}$$
be the probability that some polynomial-time quantum algorithm accepts
an input $x$.  Then the input $x$ can be encoded as a link $L = L(x)$
with bridge number $g$, so that
\eq{e:capprox}{p(x) \approx \frac{|\braket{L}_K|^2}{|t^{1/2}+t^{-1/2}|^{2g}.},}
where "$\approx$'' is in the $\FPTEAS$ sense.
\label{c:encode} \end{corollary}

Although \cor{c:encode} is essentially due to Freedman, Larsen, and Wang,
we describe one way to prove it, since it is relevant to our result.

\begin{proof}[Proof of \cor{c:encode}] First, $X(4)$ is always
two-dimensional and it can be interpreted as a qubit.  We can define its
computational basis simply by applying the Gram-Schmidt procedure to the
basis of planar matchings:
\begin{align}
\ket{0} &= \frac1{t^{1/2}+t^{-1/2}}\biggl|\;
\begin{tikzpicture}[style=basic,baseline=.5cm]
\draw (0,0) arc (-90:90:.5);
\draw (0,1.5) arc (-90:90:.5);
\end{tikzpicture}\;\biggr\rangle_K \label{e:ket0} \\
\ket{1} &= \frac{1}{\sqrt{t+1+t^{-1}}} \left(
\biggl|\;\begin{tikzpicture}[style=basic,baseline=.4cm]
\draw (0,0) arc (-90:90:1);
\draw (0,.5) arc (-90:90:.5);
\end{tikzpicture}\;\biggr\rangle_K +
\frac1{t^{1/2}+t^{-1/2}}\biggl|\;
\begin{tikzpicture}[style=basic,baseline=.5cm]
\draw (0,0) arc (-90:90:.5);
\draw (0,1.5) arc (-90:90:.5);
\end{tikzpicture}\;\biggr\rangle_K \right).
\nonumber \end{align}

Second, by \thm{th:flw} and \thm{th:sk}, a quantum circuit $C$ on $n$ qubits
can be encoded to exponential tolerance as a braid $\sigma \in B_{4n}$
on $4n$ strands. Third, the amplitude $\braket{0^n|C|0^n}$ is proportional
to the Kauffman bracket of a link $L$, which is the braid $\sigma$ capped
with $2n$ U-turns at both ends:
\eq{e:skapprox}{\braket{0^n|C|0^n} \approx \frac1{(t^{1/2}+t^{-1/2})^{2n}}
\biggl\langle\;
\begin{tikzpicture}[style=basic,baseline=.6cm]
\foreach \x in {0,2}
{
    \draw (-2,\x) -- (2,\x);
    \draw (-2,\x+1) -- (2,\x+1);
    \draw (-2,\x+1) arc (90:270:.5);
    \draw (2,\x) arc (-90:90:.5);
}
\foreach \x/\y/\s in
    {-1.5/1/0, -.5/0/1, -.5/2/0, .5/1/0}
{ 
    \begin{scope}[xshift=\x cm,yshift=\y cm,style=tangle]
    \fill[white] (.02,-.05) rectangle (.98,1.05);
    \draw[double] (0,\s) .. controls (.5,\s) and (.5,1-\s) .. (1,1-\s);
    \draw[double] (0,1-\s) .. controls (.5,1-\s) and (.5,\s) .. (1,\s);
    \end{scope}
}
\draw[dashed,red!70!black] (-2,-.5) -- (-2,3.5);
\draw[dashed,red!70!black] (2,-.5) -- (2,3.5);
\draw (0,-.5) node {$\sigma$};
\end{tikzpicture}\;\biggr\rangle_K.}
A diagram of a link $L$ in this form, a braid capped with U-turns,
is called a \emph{plat diagram}; the number of U-turns at each
end, $g = 2n$ in this case, is its \emph{bridge number}.  Finally, by
equation \eqref{e:circuit}, we can express the acceptance probability
as $|\braket{0^n|C|0^n}|^2$ where $C$ has $n = \poly(|x|)$ qubits and
$\poly(|x|)$ gates and can be generated in deterministic polynomial time
from $x$.  Combining equations \eqref{e:skapprox} and \eqref{e:circuit},
we obtain \eqref{e:capprox}, as desired.
\end{proof}

\begin{remark} In the proof of \cor{c:encode}, it is easy to worry about
leakage of amplitude into the unused part of the Hilbert space $X(4n)$.
But using the plat diagram method, \thm{th:flw} and \thm{th:sk}
applied to the unitary group $\PSU(X(8)) \cong \PSU(14)$ controls this
leakage along with the intended amplitudes.  In some other encodings of
quantum computation into the Jones polynomial, one might want a joint
denseness version of \thm{th:flw}.  It isn't needed here, although it is
needed in order to prove \thm{th:flw} itself by induction.
\end{remark}

\subsection{Proof of \thm{th:main}}
\label{s:pmain}

\begin{proof}  \cor{c:encode} describes a way to approximately ($\FPTEAS$)
encode a circuit calculation $\braket{0^n|C|0^n}$ as a plat braid with
bridge number $2g$.  This type of circuit calculation is $\BQP$-complete by
\prop{p:circuit}.  Each gate of the circuit $C$ (say a Toffoli or a Hadamard
gate, if these standard generators are used) can be approximated by a braid
by \thm{th:flw} (Freedman-Larsen-Wang) and \thm{th:sk} (Solovay-Kitaev).
Thus the left side of \eqref{e:capprox} is $\BQP$-complete in additive
approximation.   But the denominator is exponential in $g$.  This is not by
itself a hardness result, but it is a strong indication that \thm{th:ajl}
does not usually provide information about the Jones polynomial, and that
a hardness result should be available.

The first hardness result to obtain is that multiplicative approximation to
the Jones polynomial norm $|V(L,t)|$ is $\shP$-hard.  Almost by definition
(more precisely, by \prop{p:circuit}), multiplicative approximation to
the left side is $\SBQP$-hard, which by \prop{p:promise} is the same as
$\PostBQP$-hard.   The denominator on the right side is easily computable,
so we obtain that multiplicative approximation to the numerator is
also $\PostBQP$-hard.  This numerator is the Kauffman bracket value
$|\braket{L}_K|^2$, which equals $|V(L,t)|^2$, which implies hardness
of $|V(L,t)|$.   Finally, Aaronson's theorem tells us that $\PostBQP =
\PP$, and $\PP$-hard implies $\#P$-hard by \prop{p:pp}.

To complete the proof, we need to refine the construction in two ways.
We need to convert multiplicative approximation to more general
value-distinguishing approximation; and we need to change the link $L$
to a knot.

For the first refinement, let $a > b > 0$ be constants as in the statement
of \thm{th:main}, and let $p$ and $c$ be the polynomial and the constant
in the modified definition of $\SBQP$ in \lem{l:sbqp}.  By that lemma and
equation~\eqref{e:capprox}, it is $\SBQP$-complete and therefore $\shP$-hard
to determine whether
$$\frac{|\braket{L}_K|^2}{|t^{1/2} + t^{-1/2}|^{2g}}
    \begin{cases} > c2^{-p(|x|)} \\ < 2^{-p(|x|)} \end{cases}.$$
We want to make a modified link $L'$ to make it hard to determine whether
$|\braket{L'}_K|^2$ is more than $a$ or less than $b$.  Recall
that $g = \poly(|x|)$, and note that
$$|t^{1/2} + t^{-1/2}| > 1.$$
If
$$|t^{1/2} + t^{-1/2}|^{2g} \ll 2^{p(|x|)}$$
when $|x|$ is large, then we can add $m = \poly(|x|)$ copies of the unknot
to $L$ so that
$$|t^{1/2} + t^{-1/2}|^{2g+2m}2^{-p(|x|)}$$
is bounded.  On the other hand, if
$$|t^{1/2} + t^{-1/2}|^{2g} \gg 2^{p(|x|)}$$
then we can use denseness to first create a link $L_0$ (say a 2-bridge
link corresponding to a 1-qubit circuit) such that $|\braket{L_0}_K|$ is a
small constant.  Then we can add $m$ copies of $L_0$ to $L$ so that
$$|\braket{L_0}_K|^{2m}|t^{1/2} + t^{-1/2}|^{2g}2^{-p(|x|)}$$
is bounded.  The constant $c$ in the definition of $\SBQP$ can be chosen
to overwhelm the bound in either case as well as the specific values of $a$
and $b$.

Finally, we want to further modify $L'$ into a link $L''$ that has only
one component, \ie, is a knot.  The trick for this is that since the braid
group is dense, the pure braid group is also dense.  Thus we can switch
two strands, and then approximately cancel its effect with a pure braid
that does not permute any strands.  The permutation induced by the
braid is thus decoupled from the approximate value of $\braket{L''}_K$,
so $L''$ can be chosen so that it has only one component.
\end{proof}

\section{The Tutte polynomial}
\label{s:tutte}

\subsection{Tutte and Potts}
\label{s:tpotts}

In order to define
the Tutte polynomial, we will first define another graph invariant
with equivalent information known as the Potts model.  The Potts model
of a graph $G$ depends on a positive integer $q$, the number of colors;
and on a variable $y$.  The weight of a coloring of the vertices of $G$
with $q$ colors is defined as $y^k$ if $k$ of the edges of $G$ connect two
vertices of the same color.  Then the Potts partition function $Z(G,y,q)$
is defined as the total weight of all vertex colorings.  The Potts partition
function yields the Tutte polynomial $T(G,x,y)$ by the formula
$$T(G,x,y) \defeq (y-1)^{-v}(x-1)^{-c}Z(G,y,q),$$
where
\eq{e:qxy}{q = (x-1)(y-1),}
and $G$ has $v$ vertices and $c$ components.

An important variation of the Potts model (or the Tutte polynomial) is the
multivariate version, where the weight $y$ can be different for each edge
of $G$, to make a weighted graph $G(\vy)$.  Then the Potts partition function
is defined in the usual way as a multiplicative sum.  Namely, the partition
function $Z(G(\vy),q)$ is defined as the total weight of all colorings $c$
with $n$ colors; the weight of $c$ is defined as the product of the weights
$y_e$ for edges $e$ whose vertices have the same color.  Or, as a formula,
if $\cC$ is the set of colorings and $E$ is the set of edges of $G$, then
$$Z(G(\vy),q) = \sum_{c \in \cC} \prod_{\substack{(j,k) \in E \\ c(j) = c(k)}}
    y_{(j,k)}.$$
Having generalized the parameter $y$ to a weight assigned to each edge,
we still want to make use of the parameter $x$ defined from $y$ and $q$
by the relation \eqref{e:qxy}.   To this end, if we assign a weight $y$ to an
edge, we will also assign it the \emph{dual weight} $x$ using \eqref{e:qxy}.
The dual weight $x$ is simply meant as another notation for the weight $y$.
Since the dual weight $x$ is not the same number as the weight $y$, we
will denote it in the diagrams with parentheses.

The ordinary or multivariate Potts model can also be defined by a
contraction-deletion formula, together with the fact that its value for
an isolated vertex is $q$:
\begin{align}
\begin{tikzpicture}[baseline=-.75ex,style=basic]
\draw (0,0) -- (1.5,0);
\draw (-.6,-.8) -- (0,0) -- (-.6,.8);
\draw (2.1,-.8) -- (1.5,0) -- (2.1,.8);
\fill (0,0) circle (.15); \fill (1.5,0) circle (.15);
\draw (.75,0) node[anchor=south] {$y$};
\draw (-.7,0) node {.};
\draw (-.647,-.268) node {.}; \draw (-.647,.268) node {.};
\draw (2.2,0) node {.};
\draw (2.147,-.268) node {.}; \draw (2.147,.268) node {.};
\end{tikzpicture} &=
\begin{tikzpicture}[baseline=-.75ex,style=basic]
\draw (-.6,-.8) -- (0,0) -- (-.6,.8);
\draw (1.6,-.8) -- (1,0) -- (1.6,.8);
\fill (0,0) circle (.15); \fill (1,0) circle (.15);
\draw (-.7,0) node {.};
\draw (-.647,-.268) node {.}; \draw (-.647,.268) node {.};
\draw (1.7,0) node {.};
\draw (1.647,-.268) node {.}; \draw (1.647,.268) node {.};
\end{tikzpicture} + 
(y-1)\begin{tikzpicture}[baseline=-.75ex,style=basic]
\draw (-.6,-.8) -- (.6,.8);
\draw (-.6,.8) -- (.6,-.8);
\fill (0,0) circle (.15);
\draw (-.7,0) node {.};
\draw (-.647,-.268) node {.}; \draw (-.647,.268) node {.};
\draw (.7,0) node {.};
\draw (.647,-.268) node {.}; \draw (.647,.268) node {.};
\end{tikzpicture} \label{e:cd} \\
\begin{tikzpicture}[style=basic,baseline=-.8]
\useasboundingbox (-.5,-.5) rectangle (.5,.5);
\fill (0,0) circle (.15);
\end{tikzpicture} &= q. \nonumber \end{align}
(Tutte's original definition of the Tutte polynomial uses an equivalent
contraction-deletion formula.)   This second definition is important for
two reasons.   

First, it shows that the Potts partition function $Z(G,y,q)$ or $Z(G(\vy),q)$
is a polynomial in all of its parameters; it isn't only defined when $q$
is a positive integer.   Note that we can only give the Tutte polynomial
or the Potts model a complexity if each parameter such as $q$ or $y$ has
a computational complexity.  To this end, we assume that every parameter
is a real number with an $\FPTEAS$.  For no essential reason, we do not
consider complex values.

Second, the contraction-deletion formula allows us to generalize the
Potts model to a skein theory with skein spaces, in the same sense as
\sec{s:skein}.  More precisely, for each $n$ we let $F(n)_P$ be the vector
space of formal linear combinations of weighted planar graphs with $n$
marked boundary points on the outside face.  In fact, we would like to
allow some of the marked boundary points to be identical, so formally we
consider a graph $G(\vy)$ together with a function from labels to vertices,
$$f:\{1,\ldots,n\} \to V(G(\vy)),$$
which need not be either injective or surjective.   In the diagrams we
draw the boundary vertices in red.  If a vertex is marked twice or more
as a boundary, then it is drawn as multiple vertices connected by double
edges to denote that the vertices are equal.  Thus \eqref{e:cd} can be
written as follows, also using the ket notation to signify that we are
creating a skein theory.
\begin{align}
\left|\begin{tikzpicture}[baseline=-.75ex,style=basic,fill=darkred]
\useasboundingbox (-.5,-.5) rectangle (2,.5);
\draw (0,0) -- (1.5,0);
\fill (0,0) circle (.15); \fill (1.5,0) circle (.15);
\draw (.75,0) node[anchor=south] {$y$};
\end{tikzpicture}\right\rangle_P &= 
\left|\begin{tikzpicture}[baseline=-.75ex,style=basic,fill=darkred]
\useasboundingbox (-.5,-.5) rectangle (2,.5);
\fill[darkred] (0,0) circle (.15); \fill[darkred] (1.5,0) circle (.15);
\end{tikzpicture}\right\rangle_P + (y-1)
\left|\begin{tikzpicture}[baseline=-.75ex,style=basic,fill=darkred]
\useasboundingbox (-.5,-.5) rectangle (2,.5);
\draw[double,double distance=1.3] (0,0) -- (1.5,0);
\fill[darkred] (0,0) circle (.15); \fill[darkred] (1.5,0) circle (.15);
\end{tikzpicture}\right\rangle_P \label{e:pskein} \\
\left|\begin{tikzpicture}[baseline=-.75ex,style=basic]
\useasboundingbox (-.5,-.5) rectangle (.5,.5);
\fill (0,0) circle (.15);
\end{tikzpicture}\right\rangle_P &= q
\left|\begin{tikzpicture}[baseline=-.75ex,style=basic]
\useasboundingbox (-.5,-.5) rectangle (.5,.5);
\end{tikzpicture}\right\rangle_P. \nonumber
\end{align}
We then define the skein space
to be the quotient $W(n)_P = F(n)_P/\sim$, where the equivalence is given by
the relation \eqref{e:cd}.

To review, we have used \eqref{e:cd} to define skein spaces $W(n)_P$ for 
\emph{planar} graphs.  It is easy to show that one basis of $W(n)_P$
is given by noncrossing partitions of $n$ points arranged in a circle,
corresponding to graphs with no edges (other than double edges):
$$\biggl|\;\;
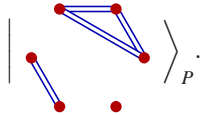
\begin{tikzpicture}[style=basic,fill=darkred,double distance=1.3,
    baseline=0]
\draw[double] (0:1.5) -- (60:1.5); \draw[double] (60:1.5) -- (120:1.5);
\draw[double] (120:1.5) -- (0:1.5); \draw[double] (180:1.5) -- (240:1.5);
\foreach \th in {0,60,120,180,240,300} \fill (\th:1.5) circle (.15);
\end{tikzpicture}\;\;\biggr\rangle_P.$$
It is well known that the number of noncrossing partitions is
the $n$th Catalan number, so that
$$\dim W(n)_P = C_n = \frac{1}{n+1}\binom{2n}{n},$$
which is the same as the dimension of the Kauffman skein space $W(2n)_K$ as
given in \eqref{e:kcat}.   In fact, the two skein theories are equivalent,
and we will make use of this coincidence to prove \thm{th:second}.

\begin{remark} What matters the most for a result such as \thm{th:second}
is that the Potts model has \emph{some} skein theory.   Although the
terminology ``skein theory" is not traditional in graph theory,
graph theorists have long used the idea of a skein theory, namely
local recurrence relations such as the contraction-deletion formula.
In particular, if we let $\tW(n)_P$ be the skein space of all graphs with
$n$ boundary vertices, not just planar graphs, then it is a standard graph
theory fact that one basis for it is the set of partitions of $n$ points.
The dimension of this skein space is the $n$th Bell number (by definition,
the number of partitions of a set with $n$ elements) rather than the $n$th
Catalan number in the planar case.
\end{remark}

\subsection{Circuits and braids}
\label{s:pcircuit}

In this section, we will define Potts quantum circuits by analogy with
the Jones braid representation and its use in the proof of \cor{c:encode}.
In particular, we will encode the standard quantum circuit evaluation
$\braket{0^n|C|0^n}$ in Potts circuit by analogy with \eqref{e:skapprox}.
Just as we did in \sec{s:skein}, we define $W(n)_P$ using graphs in the
right half-plane and we denote elements as kets $\ket{\psi}$; we define
$W'(n)_P$ using graphs in the left half-plane and we denote its elements
as bras $\bra{\psi}$.   However, we will not define any Hilbert space
structures on our skein spaces.  Instead, we will just use vector spaces
and interpret them using the libit or linear computation model defined
at the end of \sec{s:postsel}.   For concreteness, we define the initial
state $\ket{\psi} \in W(n)_P$ to be $n$ disconnected dots, and the final
state $\bra{\psi} \in W'(n)_P$ to also be $n$ disconnected dots.

Having defined initial and final states for Potts circuits, we still need to
define the circuits themselves.   We could define a Potts circuits to be any
planar graph with left and right boundary vertices.  This is the more
general possible choice; but we will define more specific quantum
gate operators $P(y)$, the parallel gate, and $S(x)$, the series gate.
A gate $P(y)$ is an edge with weight $y$, whose two
vertices are both input vertices and output vertices.  A gate $S(x)$ is an
edge with dual weight $x$ that connects an input vertex to an output vertex.
If there are $n$ vertices, then there are $n-1$ positions for $P(y)$
and $n$ positions for the gate $S(x)$; we number them
$P(y)_j$ and $S(x)_j$ starting with $j=1$.  For example, if $n=4$, then:
$$P(y)_1 = 
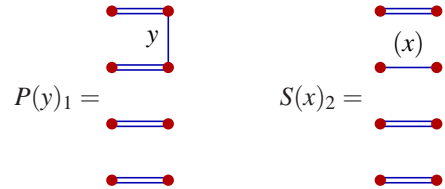
\begin{tikzpicture}[baseline=1cm,style=basic,fill=darkred]
\draw (1.5,3) -- (1.5,4.5); \draw (1.5,3.75) node[anchor=east] {$y$};
\foreach \y in {0,1.5,3,4.5}
    \draw[double,double distance=1.3] (0,\y) -- (1.5,\y);
\foreach \y in {0,1.5,3,4.5} 
    \foreach \x in {0,1.5} \fill (\x,\y) circle (.15);
\end{tikzpicture} \qquad \qquad 
S(x)_2 =\; \begin{tikzpicture}[baseline=1cm,style=basic,fill=darkred]
\draw (0,3) -- (1.5,3);
\draw (.75,3) node[anchor=south] {$(x)$};
\foreach \y in {0,1.5,4.5}
    \draw[double,double distance=1.3] (0,\y) -- (1.5,\y);
\foreach \y in {0,1.5,3,4.5} 
    \foreach \x in {0,1.5} \fill (\x,\y) circle (.15);
\end{tikzpicture}$$ 

As an example of the full circuit construction, if $n=4$,
we can make a graph $G$ composed of 8 gates so that
\begin{multline*}
Z(G(\vy);q) = \\ \braket{\psi| P(19)_1 P(17)_2 P(13)_3
    S(11)_1 S(7)_2 S(5)_4 P(3)_1 P(2)_2|\psi}.
\end{multline*}
In this example, the graph $G(\vy)$ is:
$$
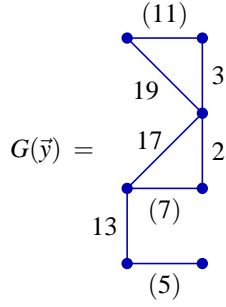
\begin{tikzpicture}[style=basic,scale=2]
\foreach \x/\y in {0/0,0/1,0/3,1/0,1/1,1/2,1/3}
    { \fill (\x,\y) circle (.075); }
\draw (0,1) -- (1,1) -- (1,3) -- (0,3) -- (1,2) -- (0,1) -- (0,0) -- (1,0);
\draw (.5,3) node[anchor=south] {$(11)$};
\draw (.5,1) node[anchor=north] {$(7)$};
\draw (.5,0) node[anchor=north] {$(5)$};
\draw (1,2.5) node[anchor=west] {$3$};
\draw (1,1.5) node[anchor=west] {$2$};
\draw (0,.5) node[anchor=east] {$13$};
\draw (.55,2.55) node[anchor=north east] {$19$};
\draw (.6,1.4) node[anchor=south east] {$17$};
\draw (-1,1.5) node {$G(\vy) \;=$};
\end{tikzpicture}.$$

To conclude this section, we show that for certain values of the parameters
$x$ and $y$, the gates $P(y)$ and $S(x)$ aren't just analogous to the
Jones braid representation; up to scalar factors, they are the Jones braid
representation.

\begin{theorem} Let $q$ and $t^{1/4}$ be parameters such that
$$q = t+2+t^{-1}.$$
Then for each $n$, there is a vector space isomorphism between the planar
Potts skein space $W(n)_P$ and the Kauffman skein space $W(2n)_K$ such
that the operators $P(-t)$ and $S(-t)$ are proportional to half-twist
generators of the Jones braid representation.
\label{th:pk} \end{theorem} 

Note that $q > 4$ in \thm{th:second}, the corresponding value of $t$
is real and positive in \thm{th:pk}, and we can also take $t^{1/4}$ to
be real and positive.  Thus, in our use of \thm{th:pk}, we can do all
calculations over the field $\R$.

\thm{th:pk} and its proof are a version of one of the earliest
constructions of the Jones polynomial of a link $L$, as the Potts partition
function of an associated graph $G(\vy)$ \cite[\S2]{Jones:pacific}.
First, the diagram of $L$ should be
given a checkerboard coloring:
$$
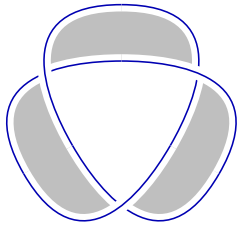
\begin{tikzpicture}[baseline=-.75ex]
\begin{scope}[style=tangle,scale=.75]
\fill[even odd rule,lightgray] (90:1)
    .. controls (135:1.414) and (165:2.828) .. (210:2)
    .. controls (255:2.828) and (285:1.414) .. (330:1)
    .. controls (15:1.414) and (45:2.828) .. (90:2)
    .. controls (135:2.828) and (165:1.414) .. (210:1)
    .. controls (255:1.414) and (285:2.828) .. (330:2)
    .. controls (15:2.828) and (45:1.414) .. (90:1);
\draw[double] (90:1) .. controls (135:1.414) and (165:2.828) .. (210:2);
\draw[double] (330:1) .. controls (15:1.414) and (45:2.828) .. (90:2);
\draw[double] (210:1) .. controls (255:1.414) and (285:2.828) .. (330:2);
\draw[double] (210:2) .. controls (255:2.828) and (285:1.414) .. (330:1);
\draw[double] (90:2) .. controls (135:2.828) and (165:1.414) .. (210:1);
\draw[double] (330:2) .. controls (15:2.828) and (45:1.414) .. (90:1);
\end{scope}
\end{tikzpicture}$$

Then we can make a weighted graph $G(\vy)$ by replacing the gray regions by
vertices, and the crossings by edges.  There are two types of crossings,
checkerboard-positive and checkerboard-negative, and they can be replaced
by edges with weight $y = -t^{\pm 1}$ (and therefore dual weight $x =
-t^{\mp 1}$):
$$\begin{array}{c@{\qquad\qquad}c}
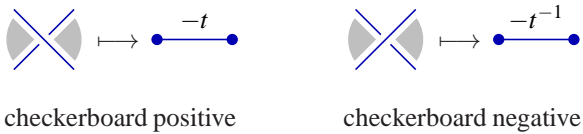
\begin{tikzpicture}[baseline=-.75ex]
\begin{scope}[style=basic]
\fill[lightgray] (0,0) -- (-.707,.707) arc (135:225:1) -- cycle;
\fill[lightgray] (0,0) -- (.707,-.707) arc (-45:45:1) -- cycle;
\end{scope}
\begin{scope}[style=tangle]
\draw[double] (-.4,-.4) -- (.4,.4); \draw[double] (-.4,.4) -- (.4,-.4);
\end{scope}
\end{tikzpicture} \;\longmapsto\;
\begin{tikzpicture}[baseline=-.75ex,style=basic]
\draw (0,0) -- (2,0);
\fill (0,0) circle (.15); \fill (2,0) circle (.15);
\draw (1,0) node[anchor=south] {$-t$};
\end{tikzpicture} &
\begin{tikzpicture}[baseline=-.75ex]
\begin{scope}[style=basic]
\fill[lightgray] (0,0) -- (-.707,.707) arc (135:225:1) -- cycle;
\fill[lightgray] (0,0) -- (.707,-.707) arc (-45:45:1) -- cycle;
\end{scope}
\begin{scope}[style=tangle]
\draw[double] (-.4,.4) -- (.4,-.4); \draw[double] (-.4,-.4) -- (.4,.4);
\end{scope}
\end{tikzpicture} \;\longmapsto\;
\begin{tikzpicture}[baseline=-.75ex,style=basic]
\draw (0,0) -- (2,0);
\fill (0,0) circle (.15); \fill (2,0) circle (.15);
\draw (1,0) node[anchor=south] {$-t^{-1}$};
\end{tikzpicture} \\[4ex]
\mbox{checkerboard positive} & \mbox{checkerboard negative} \end{array}$$
It turns out that
$$\braket{L}_K = t^{u/4} (-t^{1/2}-t^{-1/2})^{-v} Z(G(\vy),q),$$
where $u$ is the number of checkerboard-positive crossings minus the number
of checkerboard-negative crossings, and $v$ is the number of black regions
of $L$.

\begin{proof}[Proof of \thm{th:pk}] There is an evident bijection
between non-crossing partitions of $n$ points and planar matchings
of $2n$ points.   Each part of the partition is represented by a polygon
with some $k$ sides, and we can replace it by $k$ arcs:
$$
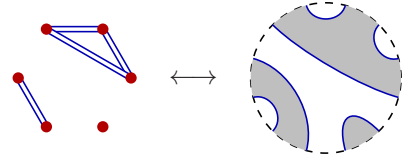
\begin{tikzpicture}[style=basic,fill=darkred,double distance=1.3,
    baseline=-.5ex]
\draw[double] (0:1.5) -- (60:1.5); \draw[double] (60:1.5) -- (120:1.5);
\draw[double] (120:1.5) -- (0:1.5); \draw[double] (180:1.5) -- (240:1.5);
\foreach \th in {0,60,120,180,240,300} \fill (\th:1.5) circle (.15);
\end{tikzpicture} \quad \longleftrightarrow \quad 
\begin{tikzpicture}[style=basic,fill=lightgray,baseline=-.5ex]
\draw[black,dashed] (0,0) circle (2);
\fill (15:2) .. controls (15:1.5) and (45:1.5) .. (45:2)
    arc (45:75:2) .. controls (75:1.5) and (105:1.5) .. (105:2)
    arc (105:135:2) .. controls (135:1) and (345:1) .. (345:2)
    arc (-15:15:2) -- cycle;
\fill (255:2) .. controls (255:1) and (165:1) .. (165:2)
    arc (165:195:2) .. controls (195:1.5) and (225:1.5) .. (225:2)
    arc (225:255:2) -- cycle;
\fill (315:2) .. controls (315:1) and (285:1) .. (285:2)
    arc (285:315:2) -- cycle;
\draw (15:2) .. controls (15:1.5) and (45:1.5) .. (45:2);
\draw (75:2) .. controls (75:1.5) and (105:1.5) .. (105:2);
\draw (135:2) .. controls (135:1) and (345:1) .. (345:2);
\draw (165:2) .. controls (165:1) and (255:1) .. (255:2);
\draw (195:2) .. controls (195:1.5) and (225:1.5) .. (225:2);
\draw (285:2) .. controls (285:1) and (315:1) .. (315:2);
\end{tikzpicture}
$$
We will use the same symbol $m$ to denote either the partition or its
corresponding matching.   The vectors $\ket{m}_P$ are a basis of
$W(n)_P$, while the vectors $\ket{m}_K$ are a basis of $W(2n)_K$.
We identify them using the formula
$$\ket{m}_P = (-t^{1/2}-t^{-1/2})^{c(m)}\ket{m}_K,$$
where $c(m)$ is the number of components of $m$ as a partition,
or the number of black regions of $m$ read as a planar matching.

With this choice of isomorphism, we claim that
if $R_j$ is the $j$th left half-twist operator on $W(n)_K$ in the 
Jones braid representation, then
\begin{align*}
S(-t)_j &= (t^{1/4} + t^{-3/4}) R_{2j-1} \\
P(-t)_j &= -t^{1/4} R_{2j}.
\end{align*}
The first of these relations is established as follows.
We do the calculation in terms of kets; the reader can
check that it works the same way with operators.   We obtain:
\begin{align*}
S(-t) &= \left|\begin{tikzpicture}[baseline=-.75ex,style=basic,fill=darkred]
\useasboundingbox (-.5,-.5) rectangle (2,.5);
\draw (0,0) -- (1.5,0);
\fill (0,0) circle (.15); \fill (1.5,0) circle (.15);
\draw (.75,0) node[anchor=south] {$(-t)$};
\end{tikzpicture}\right\rangle_P = 
\left|\begin{tikzpicture}[baseline=-.75ex,style=basic,fill=darkred]
\useasboundingbox (-.5,-.5) rectangle (2,.5);
\draw (0,0) -- (1.5,0);
\fill (0,0) circle (.15); \fill (1.5,0) circle (.15);
\draw (.75,0) node[anchor=south] {$-t^{-1}$};
\end{tikzpicture}\right\rangle_P \\
&= \left|\begin{tikzpicture}[baseline=-.75ex,style=basic,fill=darkred]
\useasboundingbox (-.5,-.5) rectangle (2,.5);
\fill[darkred] (0,0) circle (.15); \fill[darkred] (1.5,0) circle (.15);
\end{tikzpicture}\right\rangle_P - (1+t^{-1})
\left|\begin{tikzpicture}[baseline=-.75ex,style=basic,fill=darkred]
\useasboundingbox (-.5,-.5) rectangle (2,.5);
\draw[double,double distance=1.3] (0,0) -- (1.5,0);
\fill[darkred] (0,0) circle (.15); \fill[darkred] (1.5,0) circle (.15);
\end{tikzpicture}\right\rangle_P \\
&= -(t^{1/2} + t^{-1/2})
\left|\begin{tikzpicture}[baseline=-.75ex,style=basic]
\useasboundingbox (-.6,.6) rectangle (.6,.6);
\draw (-.5,-.5) arc (-45:45:.707);
\draw (.5,.5) arc (135:225:.707);
\end{tikzpicture}\right\rangle_K
- (1+t^{-1}) \left|\begin{tikzpicture}[baseline=-.75ex,style=basic]
\useasboundingbox (-.6,.6) rectangle (.6,.6);
\draw (.5,-.5) arc (45:135:.707);
\draw (-.5,.5) arc (225:315:.707);
\end{tikzpicture}\right\rangle_K \\
&= (t^{1/4} + t^{-3/4})
\left|\begin{tikzpicture}[baseline=-.75ex,style=stangle]
\useasboundingbox (-.6,.6) rectangle (.6,.6);
\draw[double] (-.5,.5) -- (.5,-.5);
\draw[double] (-.5,-.5) -- (.5,.5);
\end{tikzpicture}\right\rangle_K
\end{align*}
using \eqref{e:pskein} and \eqref{e:skein}.
(The extra factor of $- t^{1/2} - t^{-1/2}$ in the first term
arises from the change of basis from Potts skeins to Kauffman skeins.)
The calculation for $P(-t)$ is similar.

Since the braid generators are proportional to the parallel and series
operators, the latter generate the same projective representation.
\end{proof}

\subsection{Parallel-series compositions}
\label{s:comps}

The statement of \thm{th:second} only allows graphs with the same weight $y$
for every edge.   If we want to use the gates $P(y)$ and $S(x)$ universal
quantum computation, this is not even enough for the Solovay-Kitaev theorem,
if we don't have the inverses of these two gates.   In this section we
use a technique used by Goldberg and Jerrum in which edges are replaced
by subgraph gadgets, to approximately allow any real weight $y$ for any
edge \cite{GJ:tutte}. This will give let use the Solovay-Kitaev theorem
by the relations
$$P(y)^{-1} = P(y^{-1}) \qquad S(x)^{-1} \propto S(x^{-1}),$$
which follow from \eqref{e:psop} below.  It will also
make it easier to prove the dense generation criterion that is
also needed for the Solovay-Kitaev theorem.

The technique is as follows:  If a graph $G(\vy)$ has two parallel edges
with weight $y_1$ and $y_2$, then they are equivalent to a single edge with
weight $y_1y_2$.  Meanwhile, if $G(\vy)$ has two edges in series with dual
weight $x_1$ and $x_2$, they are equivalent (up to changing the Potts value
$Z$ by a constant factor) to one edge with weight $x_1x_2$.  In other words,
\eq{e:psop}{P(y_1y_2) = P(y_1)P(y_2) \qquad S(x_1x_2) \propto S(x_1)S(x_2).}
These transformations are called \emph{shift operations}; they are also
called \emph{compositions} and \emph{implemented weights}.  Note that
series and parallel compositions preserve the value of $q$, and they
preserve planarity.

\begin{lemma} Consider graphs with the Potts model with $q$ colors
and with a single weight $y$ which is an $\FPTEAS$ number.  Suppose that
$q > 4$ and that $x,y < 0$.  Then all weights $y' \ne 1$ that are $\FPTEAS$
numbers, can be $\FPTEAS$ approximated by parallel and series compositions.
\label{l:ally} \end{lemma}

\lem{l:ally} is a refinement of one proved by Goldberg and Jerrum
\cite{GJ:tutte}.  (The refinement is that they did not establish is the
$\FPTEAS$ property.)

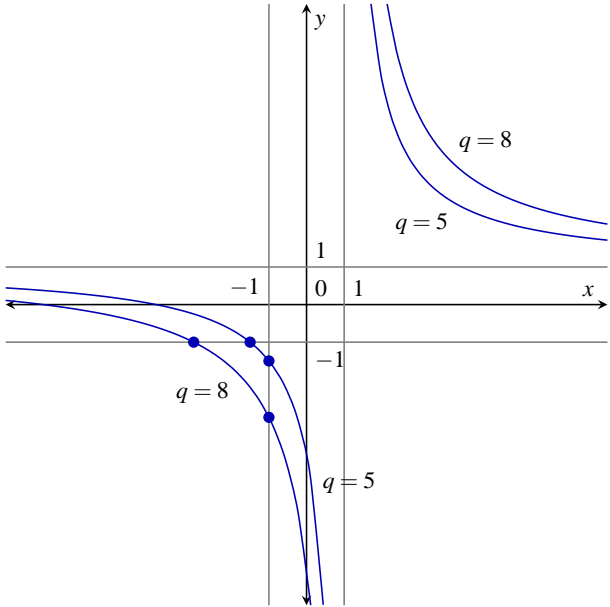
\begin{fullfigure}{f:tutteplane}{The Tutte plane with level curves of $q$}
\begin{tikzpicture}[style=basic]
\draw[black,<->] (-8,0) -- (8,0); \draw[black,<->] (0,-8) -- (0,8);
\draw[gray] (-8,-1) -- (8,-1); \draw[gray] (-8,1) -- (8,1);
\draw[gray] (-1,-8) -- (-1,8); \draw[gray] (1,-8) -- (1,8);
\draw[anchor=south east] (-1,0) node {$-1$};
\draw[anchor=south west] (1,0) node {1};
\draw[anchor=north west] (0,-1) node {$-1$};
\draw[anchor=south west] (0,0) node {0};
\draw[anchor=south west] (0,1) node {1};
\draw[anchor=south] (7.5,0) node {$x$};
\draw[anchor=west] (0,7.5) node {$y$};
\draw[anchor=north east] (4,2.667) node {$q=5$};
\draw[anchor=south west] (.2,-5.25) node {$q=5$};
\draw[anchor=south west] (3.828,3.828) node {$q=8$};
\draw[anchor=north east] (-1.828,-1.828) node {$q=8$};
\draw[smooth,domain=-8:0.444] plot (\x,{5/(\x-1)+1});
\draw[smooth,domain=1.714:8] plot (\x,{5/(\x-1)+1});
\draw[smooth,domain=-8:0.111] plot (\x,{8/(\x-1)+1});
\draw[smooth,domain=2.143:8] plot (\x,{8/(\x-1)+1});
\foreach \x/\y in {-1/-1.5,-1.5/-1,-3/-1,-1/-3} { \fill (\x,\y) circle (.15); }
\end{tikzpicture} \end{fullfigure} 

\begin{proof} \fig{f:tutteplane} shows a diagram of curves in the $x$-$y$
plane (the Tutte plane) with constant values of $q$.  Given that $q > 4$
and $x,y < 0$, we must have either that $x < -1$ or $y < -1$ or both.
Parallel composition has the same effect on $y$ as series composition
has on $x$, and vice versa; so we can assume without loss of generality
that $x < -1$.  As a first step, we can create the dual weight $x^n$ with
a series composition with $n$ edges.  This creates a sequence of weights
$y_n$ that satisfies the estimate
$$\log(y_n) = qx^{-n}(1+o(1))$$
as $n \to \infty$.  Now suppose that $y' > 1$ is some other weight.
We claim that we can efficiently approximate $y'$ as a product
of weights $y_{2n}$.  Equivalently, we claim that
we can efficiently approximate $\log(y')$ as a sum of terms $\log(y_{2n})$:
$$\log(y') = \log(y_{2n_1}) + \log(y_{2n_2}) + \cdots.$$
This can be viewed as a bin packing problem, because both $\log(y')$ and
each term $\log(y_{2n})$ are positive.  The claim is established by using
a greedy bin-packing algorithm.  \Ie, choose each term $\log(y_{2n_k})$
to be as large as possible, but so that the partial sum does not exceed
$\log(y')$.  Since the terms $\log(y_{2n})$ decrease exponentially (and no
faster), and since the graph complexity of each term is linear in $n$,
the result is a parallel-series composition which is an $\FPTEAS$ for the
weight $y'$.

The same bin-packing argument works for $0 < y' < 1$, using the
odd-numbered weights $y_{2n+1}$.  So every desired weight $y' > 0$ has
an $\FPTEAS$-strength parallel-series composition.  In addition, we also
have the original weight $y < 0$, so the values of $y' > 0$, $y' = y''y$
with $y'' > 0$, and $y$ itself reach every desired value other than $y'
= 0$.  Since we also want the remaining weight $y' = 0$, we can at this
point achieve its dual weight $x' = 1-q$ with a series composition with
the dual weights $x' = -1$ and $x' = q-1$.
\end{proof}

\subsection{Densely generating $\PSL(W(n)_P)$}
\label{s:dense}

In this section, we will prove that if $q > 4$, then there are $\FPTEAS$
numbers $x$, $y_1$, and $y_2$, such that the gates $S(x)$, $P(y_1)$,
and $P(y_2)$ and their inverses densely generate the group $\PSL(W(n)_P)$
for any $n \ge 2$.   \lem{l:ally} says that we can obtain any such gates
in $\FPTEAS$ approximation using subgraph gadgets.  Our argument borrows
from the author's previous work \cite{Kuperberg:zdense} and makes crucial
use of the Zariski topology on the group $\PSL(W(n)_P)$.

The \emph{Zariski topology} on an algebraic group (or any algebraic variety)
is by definition the topology in which the closed sets are solutions to
polynomial equations.  The Zariski topology on $\R^n$ or on $\PSL(n,\R)$
is much coarser than the standard topology, which in this context is called
the \emph{analytic topology}.   It is easier for a subgroup or a subset
to be Zariski dense, and it is easier to prove Zariski denseness in this
algebraically adapted topology.   In particular:

\begin{theorem} \cite[Cor. 1.2]{Kuperberg:zdense} Let $n > 1$ be an integer
and let $t > 1$ be real.   Then the Jones braid representation of
$B_{2n}$ acting on $W(2n)_K = X(2n)_K$ with parameter $t$ is 
Zariski dense in $\PSL(X(2n))$.
\label{th:zdense} \end{theorem}

On the other hand, in some circumstances we can get the best of both worlds:

\begin{proposition} \cite[\S3]{Kuperberg:zdense} A subgroup $\Gamma$
of a connected, simple Lie group $G$ is analytically dense if and only if
it is both analytically indiscrete and Zariski dense.
\label{p:adense} \end{proposition}

(\prop{p:adense} is a baby version of a more famous result known as the
Zassenhaus neighborhood theorem \cite{Zassenhaus:diskrete,Kapovich:book}.)

To finish the construction, let $q = t+2+t^{-1}$, let $x = y_1 = -t$ and
$y_2 = t^{\sqrt{2}}$.  (The the only requirement is that $y_2$ should be
an irrational power of $t$ with an $\FPTEAS$ exponent.)  Then the gates
$P(y_1)$ and $P(y_2)$ generate an indiscrete group by \eqref{e:psop};
their products
$$P(-t)^aP(t^{\sqrt{2}})^b = P((-1)^at^{a+\sqrt{2}b})$$
for all $a,b \in \Z$ are a dense subset of all $P(y)$.  By \thm{th:pk},
the gates $S(x)$ and $P(y_1)$ acting on $W(n)_P = W(2n)_K$ generate the
Jones braid representation of $B_{2n}$.   By \thm{th:zdense}, this group
action is Zariski dense.  With the addition of the gate $P(y_2)$, it is
also indiscrete and therefore analytically dense by \prop{p:adense}.

\begin{remark} A self-contained proof of \thm{th:second} would be simpler
if we applied some of the techniques involved in \thm{th:zdense} directly
to the group generated by gates of the form $P(y)$ and $S(x)$.  However,
these techniques involve yet another set of mathematical tools that we
prefer to relegate to \cite{Kuperberg:zdense}.
\end{remark}

\subsection{Proof of \thm{th:second}}
\label{s:psecond}

\begin{proof} Following \cor{c:encode} and its proof, let
$$p(x) > 2^{-\poly(|x|)}$$
be the probability that some polynomial-time quantum
algorithm accepts an input $x$.
Then 
$$p(x) = |\braket{0^n|C(x)|0^n}|^2$$
for some a quantum circuit $C(x)$ that can be generated from $x$ in
(classical) polynomial time.   We can use the 2-dimensional skein space
$W(2)_P$ as a libit, and let $\ket{0} = \ket{\psi}$ be the state of
two dots as in \sec{s:pcircuit}.   By \lem{l:ally}, we can approximate
the gates $S(-t)$, $P(-t)$, and $P(t^{\sqrt{2}})$ and their inverses.
By \sec{s:dense}, these gates densely generate $\PSL(W(4)_P) \cong
\PSL(14,\R)$ in the case $n=4$.  Then we can apply Solovay-Kitaev,
\thm{th:sk}, to approximately encode the gates of $C(x)$ as a circuit
acting on $\PSL(W(2n)_P)$.  Then we finalize the circuit with the states
$\bra{0}$, which can also be defined as the state $\bra{\psi}$ of two dots.

The result is a graph $G(x)$ such that the Potts value $Z(G(x),q,y)$
satisfies
$$p(x) \approx N(x)|Z(G(x),q,y)|^2,$$
where the extra factor $N(x)$ is a polynomial-time computable normalization
that depends on the construction of $G(x)$.   (The factor of $N(x)$
appears because we are working up a scalar factor in all of our computations.
Note also that the $x$ here is the decision problem input and not the Potts 
parameter.)  It follows that for every $c > 1$, multiplicative approximation
of $Z(G(x),q,y)$ up to a factor of $c$ is $\PostBQP$-hard, and thus
$\shP$-hard. \end{proof}

\section{Final remarks and questions}
\label{s:final}

\subsection{Other properties of knots}

\thm{th:main} says that value-distinguishing approximation of certain values
of the Jones polynomial are $\shP$-hard even when the link $L$ is taken
to be a knot.  We conjecture that $L$ could in addition be a prime knot
or even an atoroidal knot.  (A \emph{prime} knot is one which is not a
composite of two knots; an \emph{atoroidal} knot is one which is not a
satellite \cite[\S2.C]{BZ:knots}.)  Maybe other such restrictions on
the structure of $L$ could be imposed.  But without a result such as
that distinguishing the unknot (say) is hard, it is not feasible to add
arbitrary interesting topological restrictions on $L$ to \thm{th:main}.
Maybe recognizing the unknot is in $\P$ or $\BQP$.  The Jones polynomial
would then be easy to compute for knots that are recognized as the unknot
or recognized as some other specific knots.

In fact, recognizing the unknot is in $\NP$ \cite{HLP:complexity},
and in $\coNP$ assuming the generalized Riemann hypothesis
\cite{Kuperberg:knottedness}.  Thus, unless the polynomial hierarchy
collapses, recognizing the unknot has lower qualitative computational
complexity than approximating the Jones polynomial.  (But the Jones
polynomial could still have competitive \emph{quantitative} complexity,
\ie, asymptotic time complexity in a realistic computational model.)

\subsection{Other kinds of approximation}

There are many other kinds of partial information about the Jones polynomial
without any interesting complexity bound to our knowledge.  Is the degree
of the Jones polynomial intractable?  Is it intractable to determine when
some value of the Jones polynomial vanishes?  What if the Jones polynomial
is reduced mod $p$ for some prime $p$?

\subsection{Denseness may be more than necessary}

It is easiest to see that a set of gates is universal for linear computation
if they densely generate an appropriate Lie group.  For instance, they
might generate $\PSL(2^n,\C)$ if they act on $n$ libits, or $\PSL(2^n,\R)$
inside it.  But dense generation is more than necessary for certain
types of universality.  For example, $k$-libit gates with integer matrices
always generate a discrete group, even when acting on $n > k$ libits.
Nonetheless, both the Hadamard and Toffoli gates are proportional to
integer gates, and they are universal for quantum computation.  Thus,
multiplicative approximation of amplitudes in linear computation with
integer gates is $\shP$-hard.  We do not know the right criteria on linear
gates to establish $\shP$-hardness results.

\subsection{Solovay-Kitaev without inverses}

It is a long-standing open problem to generalize the celebrated
Solovay-Kitaev theorem to gate sets that are not closed under inverses.
This problem could be peripheral in the context of designing actual quantum
computers or realistic quantum algorithms.  However, it could be important
for the purpose of establishing hardness results.

\subsection{Morse algorithms may be optimal}

It is common practice to compute the Jones polynomial by a strategy
known variously as a Morse algorithm, dynamic programming, a scanline
algorithm, or a divide-and-conquer algorithm.  (Morse theory in geometric
topology is a theory of analyzing a topological object by dividing it into
horizontal slices.)  For a knot in a plat diagram, the strategy is
to numerically compute the action of the braid group on the skein space.
This type of algorithm requires simple exponential time and space in
the number of strands of the braid, or for other kinds of knot diagrams,
the width of the diagram.  This is much better than a direct recursive
evaluation of the Jones polynomial using a finite set of skein relations;
the time complexity of any such direct algorithm is instead exponential
in the number of crossings.

It is natural to wonder whether there are other clever algorithms that
can compute the Jones polynomial even faster.  The proof of \thm{th:main}
could be evidence that Morse algorithms are essentially optimal for many
kinds of knot diagrams.  In short, if braids are evaluated using the Jones
polynomial at the dense roots of unity of \thm{th:main}, then they are a
model of general planar quantum circuits.

In more detail, consider a typical hard search problem based on classical
circuits, and an analogous problem based on quantum circuits.  For instance,
let $(z,w) = C(x,y)$ be a reversible circuit whose input $(x,y)$ and output
$(z,w)$ are each divided into two registers of equal length.  Then it is
$\NP$-hard to determine whether there is a solution to $(z,0) = C(x,0)$.
We conjecture that there are linear-depth, planar circuits $C$ for which
this problem requires exponential time in $|x|$, in other words that full
cryptography can be achieved with linear depth, planar circuits.

Using denseness at a non-lattice root of unity and Solovay-Kitaev,
\thm{th:sk}, this circuit problem can be encoded in a braid with polynomial
overhead.  (Again, the Solovay-Kitaev theorem has polylogarithmic overhead
for $\BQP$, but polynomial overhead for $\PostBQP$.)  We conjecture that
this extra polynomial overhead is not essential for hardness.  We have
in mind that there could be cryptographic methods to make linear-depth
plat diagrams of knots, for which the Jones polynomial requires
exponential time in the bridge number $g$ to estimate at a non-lattice root
of unity.  (Note that the depth of a braid is not the same as its length;
to calculate the depth, commuting half-twists can be applied in parallel.)
Such conjectures are very difficult to prove unconditionally, because they
would imply that $\shP$ is not contained in $\FP$.  Nonetheless, if there
were a believable theory of cryptography for the Jones representation of
linear-depth braids, then one would also believe that Morse algorithms
to compute or estimate the Jones polynomial are essentially optimal.


\providecommand{\bysame}{\leavevmode\hbox to3em{\hrulefill}\thinspace}
\providecommand{\MR}{\relax\ifhmode\unskip\space\fi MR }
\providecommand{\MRhref}[2]{%
  \href{http://www.ams.org/mathscinet-getitem?mr=#1}{#2}
}
\providecommand{\href}[2]{#2}
\providecommand{\eprint}{\begingroup \urlstyle{tt}\Url}

\end{document}